\documentclass[a4paper,11pt]{article}

\usepackage{jinstpub} 
\usepackage{lineno}
\usepackage{booktabs}
\usepackage{multirow}
\usepackage{tabularx}
\usepackage{subcaption}
\usepackage{amsthm}

\usepackage[compatibility=false]{caption}
\usepackage{makecell}
\usepackage{orcidlink}

\makeatletter
\gdef\@fpheader{Published in \href{https://doi.org/10.1088/1748-0221/21/07/P07037}{JINST 21 {\upshape(2026)} P07037}}
\makeatother

\usepackage{xcolor}

\usepackage[
    separate-uncertainty=true, 
    round-mode=uncertainty,
    round-precision=1,
]{siunitx}

\title{\boldmath Cross-Geometry Transfer Learning in Fast Electromagnetic Shower Simulation}

\author[b]{Frank Gaede\,\orcidlink{0000-0002-7055-9200},}

\author[a]{Gregor Kasieczka\,\orcidlink{0000-0003-3457-2755},}

\author[a,1]{Lorenzo Valente\,\orcidlink{0009-0007-0080-8738}\note{Corresponding author. Authorship is alphabetical.}}
\emailAdd{lorenzo.valente@uni-hamburg.de}

\affiliation[a]{Institut für Experimentalphysik, Universität Hamburg,\\
Luruper Chaussee 149, 22607 Hamburg, Germany}
\affiliation[b]{Deutsches Elektronen-Synchrotron DESY,\\
Notkestr. 85, 22607 Hamburg, Germany}

\abstract{Accurate particle shower simulation remains a critical computational bottleneck for high-energy physics.
Traditional Monte Carlo methods, such as \textsc{Geant4}, are computationally prohibitive, while existing machine learning surrogates are tied to specific detector geometries and require complete retraining for each design change or alternative detector. 
We present a transfer learning methodology for generative calorimeter simulation models that enables adaptation across diverse geometries with high data efficiency. 
Using point cloud representations and pre-training on the International Large Detector, our approach handles new configurations without re-voxelizing showers for each geometry. 
On the \textsc{CaloChallenge} dataset, transfer learning with only 100 target-domain samples achieves a $51\%$ improvement on the geometric mean of Wasserstein distance over training from scratch. 
Parameter-efficient fine-tuning with bias-only adaptation achieves competitive performance while updating only $17\%$ of model parameters. 
Our analysis provides insight into adaptation mechanisms for particle shower development, establishing a baseline for future progress of point cloud approaches in calorimeter simulation.
}


\arxivnumber{2512.00187}

\begin{document}
\maketitle
\flushbottom

\newpage
\section{Introduction}
\label{sec:intro}

The next decade of large-scale experiments in high-energy physics (HEP) will produce experimental data at unprecedented volumes.
This increase is driven by the higher collision rates expected at the High-Luminosity Large Hadron Collider (HL-LHC) and by the deployment of high-granularity detectors with an expanding number of readout channels \cite{CERN-LHCC-2022-005}. 
While \textsc{Geant4} \cite{GEANT4:2002zbu} provides accurate physics simulation, a single HL-LHC event may require minutes of CPU time to simulate \cite{Gavranovic:2023oam}.
In particular, calorimeter shower development constitutes the dominant computational bottleneck in detector simulation \cite{Software:2815292, ATLAS:1300517, ATL-SOFT-PUB-2014-001, Abdullin:2011zz, Hildreth_2017vpw}.
This growing computational demand cannot be satisfied solely by hardware improvements.
Single-core CPU performance has essentially plateaued: Moore's Law scaling no longer delivers the improvements we need \cite{HEPSoftwareFoundation:2017ggl}, making fundamental algorithmic innovations essential rather than relying on incremental optimisations.

The HEP community has looked to machine learning as a potential acceleration method in response to these computing limitations.
These fast simulation (FastSim) techniques learn to predict the final detector response directly from incident particle attributes instead of modelling particle interactions step-by-step through detector materials, potentially leading to orders of magnitude speedups.
Recently, significant progress has been made in the development of surrogate simulators based on generative modeling approaches \cite{Hashemi:2023rgo, Ahmad:2024dql, ATLAS:2021pzo, Barbetti:2023bvi},
ranging from generative adversarial networks \cite{Paganini:2017dwg, ATLAS:2020quw, Khattak:2021ndw, Erdmann:2018jxd, Musella:2018rdi, deOliveira:2017pjk, deOliveira:2017rwa, Paganini:2017hrr, rukh2018three, Vallecorsa:2019ked, Belayneh:2019vyx, Chekalina:2018hxi, Diefenbacher:2020rna, Jaruskova:2023cke, FaucciGiannelli:2023fow, Carminati:2018khv, Erdmann:2023ngr},  to variational auto-encoders \cite{ATL-SOFT-PUB-2018-001, Buhmann:2020pmy, Cresswell:2022tof, Diefenbacher:2023prl,  Salamani:2023ttx, Raikwar:2024peb, liu2024calo, deja2020end, Buhmann:2021lxj, Buhmann:2021caf, Hariri:2021clz, AbhishekAbhishek:2022wby}, flow-based models \cite{Krause:2021ilc, Krause:2021wez, Diefenbacher:2023vsw, Ernst:2023qvn, Buss:2024orz, Krause:2022jna, Pang:2023wfx, Dreyer:2024bhs, Erdmann:2025tsq, Smith:2024lxz}, diffusion models \cite{Mikuni:2022xry, Amram:2023onf, Mikuni:2023tqg, Favaro:2024rle,  Diefenbacher:2023flw, Acosta:2023zik,Kobylianskii:2024ijw,Kobylianskii:2024sup} and autoregressive models \cite{Lu:2020npg, Liu:2022dem,Liu:2023lnn,Buckley:2023daw}.
While these methods have demonstrated impressive performance on standardised benchmarks \cite{calochallenge2022}, most remain tied to a specific detector geometry: when detector designs evolve, or conditions change during data taking, these models typically require complete retraining with new simulation datasets.
Recent efforts have begun addressing this limitation through meta-learning \cite{Salamani:2023ttx}, multi-geometry pre-training \cite{Raikwar:2025fky}, and geometry-parametrised architectures \cite{Liu:2022dem, Smith:2024lxz}, but the general problem of efficient cross-geometry adaptation remains open.

Point cloud representations have emerged to address geometry dependence \cite{buss2025calohadronic, Schnake:2024mip, Buhmann:2023bwk, Buhmann:2023kdg, Buss:2025kiu}, generating showers as 3D space points with associated energy depositions that can, in principle, project onto arbitrary detector configurations. 
Recent work has demonstrated that point cloud models can achieve a favourable balance between speed and accuracy for highly granular calorimeter simulation in realistic applications \cite{Buss:2025bec}, validating this representation choice for practical deployment.
While this flexibility comes with computational overhead: variable-cardinality management, sparse representations with $\mathcal{O}(10^4)$ points, and complex detector reintegration, the more fundamental challenge is that representation flexibility alone does not guarantee successful transfer. 
Cross-geometry generalisation requires both the geometric flexibility of point clouds and the learnt physics knowledge that generalises across detectors. This work investigates whether single-detector pre-training on point clouds can provide both representation flexibility and model transferability, treating them as complementary rather than equivalent capabilities.

The foundation model paradigm from Natural Language Processing (NLP) and computer vision \cite{bommasani2022opportunitiesrisksfoundationmodels, reed2022a, Brown:2020mpj} offers a natural framework for developing generalisable simulation models. 
Early explorations include \textsc{OmniJet-$\alpha$}, which introduced the first general-purpose HEP model for classification and jet generation \cite{Birk:2024knn, Amram:2024fjg}, demonstrating the feasibility of unifying multiple tasks within a single architecture. 
This was later extended to showers in \textsc{OmniJet}-$\alpha_C$ \cite{Birk:2025wai}, though it lacked any pre-training or adaptation mechanism.

 More recently, \textsc{CaloDiT-2} \cite{Raikwar:2025fky} demonstrated successful pre-training on four detector geometries from the \textsc{LEMURS} dataset \cite{McKeown:2025gtw}, achieving effective transfer through standard fine-tuning, marking a first step towards a potential FastSim foundation  \cite{Radford2018ImprovingLU}.
Our work differs from \textsc{CaloDiT-2} in two key aspects.
First, in data scope, we focus on single-detector pre-training to explore scenarios where only one well-characterized detector dataset is available for pre-training, rather than requiring multiple diverse detector datasets as in \textsc{CaloDiT-2}. 
This reflects practical constraints where comprehensive simulation data may exist for established detectors but not for new designs under development. 
Second, in representation, we employ point clouds rather than fixed grids, trading some computational overhead for geometric flexibility and a direct match to the sparse nature of calorimeter showers.
When combined with Parameter-Efficient Fine-Tuning (PEFT) \cite{pmlr-v97-houlsby19a}, this approach aims to simplify the adaptation pipeline and reduce computational requirements for model scaling.
This paper investigates the feasibility of single-detector transfer learning for point cloud calorimeter simulation, focusing on parameter-efficient adaptation strategies and the underlying physics transformations that influence transferability.

The remainder of this paper is structured as follows: 
Sec. \ref{sec:transfer_learning} details the model architecture and transfer learning methodology.
Sec. \ref{sec:datasets} describes the datasets used for pre-training and fine-tuning. 
Sec. \ref{sec:results} defines the evaluation metrics and presents cross-calorimeter transfer learning results across different fine-tuning techniques. 
Sec. \ref{sec:conclusions} concludes with discussion and future directions.

\section{Cross-Calorimeter Transfer Learning}
\label{sec:transfer_learning}
This section describes the model architecture and transfer learning methodology used to adapt calorimeter simulation across different detector geometries.

\subsection{Model Architecture}
\label{sec:architecture}

The present work uses the \textsc{CaloClouds} \cite{Buhmann:2023bwk, Buhmann:2023kdg, Buss:2025kiu} network architecture as the base model to simulate electromagnetic showers across different calorimeter geometries. 
This framework comprises two complementary generative models: 

\textbf{\textsc{PointWise Net}} employs a diffusion model following the EDM (Elucidating Design Space) framework \cite{NEURIPS2022_a98846e9} to generate the spatial coordinates $(x, y, z)$ and energy depositions $e$ of shower hits as continuous point clouds. 
A detailed description is available in Ref. \cite{Buhmann:2023kdg}.
The final layer produces the denoised point cloud prediction, which enables the generation to be independent of specific detector voxelization, allowing projection onto arbitrary geometric configurations.
While point clouds provide flexibility in the transverse plane $(x, y)$, longitudinal variations in detector materials fundamentally alter the physics of shower development through changes in radiation length and interaction properties, requiring retraining rather than simple geometric projection.

\textbf{\textsc{ShowerFlow}} predicts the number of points per calorimeter layer $N_{z,i}$ that subsequently condition \textsc{PointWise Net}'s generation process. 
The architecture uses normalising flow blocks (detailed in Appendix~\ref{app:model hyperparms}), trained to learn the relationship between incident particle energy and layer-wise shower occupancy.
For \textsc{ShowerFlow} training, we apply a fixed-scale normalisation strategy that differs from the original \textsc{CaloClouds} implementation. 
Rather than normalising each event's point counts to $[0, 1]$ independently, a constant normalisation value $\texttt{norm\_points} = 800$ is applied across all events for each calorimeter layer. 
This choice is motivated by the hypothesis that the event-wise normalisation might compress the ranges in ways that could obscure scale information relevant for transfer learning across datasets with different energy and occupancy distributions.

\subsection{Transfer Learning Framework}
\label{sec:tl_framework}

\begin{figure}[htbp]
    \centering
        \includegraphics[width=1.0\textwidth]{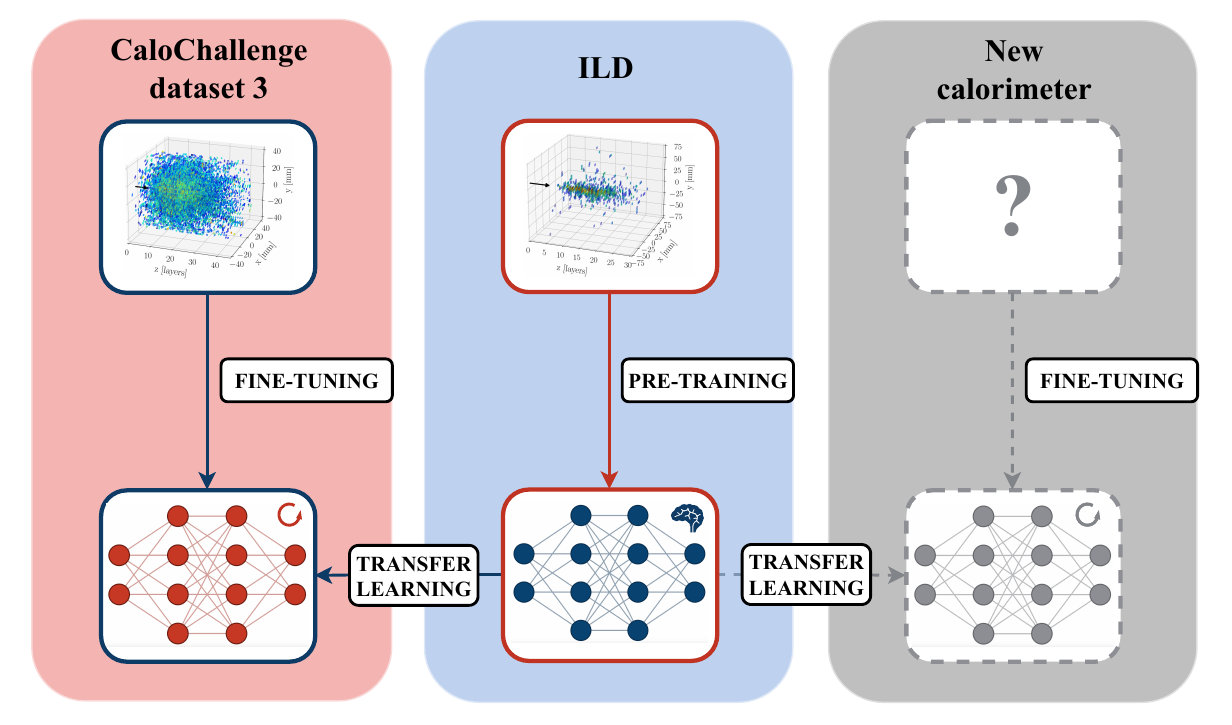} 
    \caption{ The transfer learning approach presented in this work. 
    A model pre-trained on the ILD detector is adapted to new geometries, such as \textsc{CaloChallenge} Dataset 3, through fine-tuning. 
    This approach contrasts with the conventional "from scratch" paradigm, where models are initialised with random weights and must learn all physics representations directly from the target dataset.
    The dashed box with a question mark represents potential future applications to additional detector configurations.} 
    \label{fig:overview}
\end{figure}

The approach adapts a pre-trained model, initially trained on photon-induced showers in the International Large Detector (ILD) geometry, to enable unsupervised knowledge transfer to different calorimeter configurations.
This methodology eliminates the requirement for labelled data correspondence that characterises supervised approaches in similar applications \cite{Mokhtar:2025zqs, Chappell:2022yxd, Birk:2024knn, Golling:2024abg, Li:2024htp, Mikuni:2024qsr, Kuchera:2018djs, Tombs:2021wae, Dreyer:2022yom, Beauchesne:2023vie, Bhimji:2025isp}, the conceptual approach is illustrated in Figure~\ref{fig:overview}.

We evaluate cross-geometry adaptation through two primary training strategies:

\begin{description}
    \item[\textbf{from scratch,}] in which models are initialised with random weights, representing the conventional training paradigm where each detector geometry requires complete model training. This serves as our baseline for quantifying the benefits of transfer learning.
    
    \item[\textbf{fine-tuning,}] in which models are initialised from weights pretrained on ILD photon showers, then all parameters are updated during adaptation to the \textsc{CaloChallenge} electron shower task. This tests whether learned representations generalise across different detector conditions.
\end{description}

\begin{figure}[htbp]
    \centering
    \includegraphics[width=0.99\linewidth]{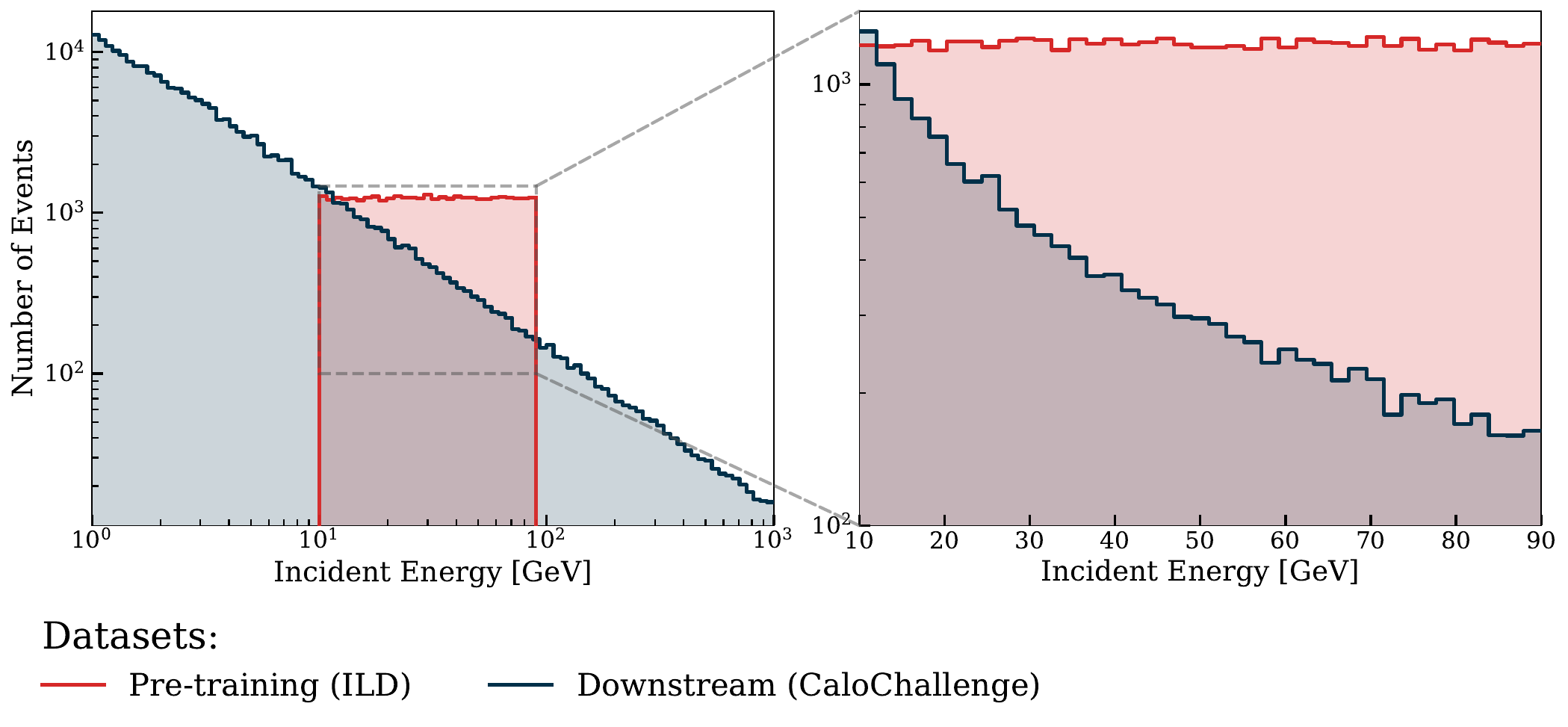}
    \caption{Incident energy distributions for pre-training (ILD, red, uniform 10-90 GeV) and downstream (CaloChallenge, blue, log-uniform 1-1000 GeV) datasets.
    Left: Full range, with a dashed box indicating the overlap region. Right: Magnified overlap showing distributional differences that, combined with particle type and geometry shifts, constitute the compound domain shift addressed in this work.}
    \label{fig:ds_ecomparison}
\end{figure}

The transfer presents multiple simultaneous challenges. 
First, the detector geometry changes from planar (\textsc{ILD}) with rectangular cells to cylindrical (\textsc{CaloChallenge}) with radial-azimuthal segmentation. 
The transfer challenge persists at $\eta=0$ where both detectors have flat layers, since \textsc{ILD} uses rectangular cells while \textsc{CaloChallenge} employs curved arc-shaped voxels in $(r, \varphi)$, fundamentally altering how generated point clouds project onto the readout structure.
Second, the readout granularity differs: $30$ layers versus $45$ layers. 
Third, the incident energy range and distributions, where the target dataset (downstream) extends beyond the pre-training uniformly distributed in $10-90$ GeV to test the extrapolation capabilities of log-uniform distributed at both low ($1-10$ GeV) and high ($90-1000$ GeV) energies, as shown in Figure \ref{fig:ds_ecomparison}. 
Fourth, the particle type changes from photons to electrons, though both produce electromagnetic cascades governed by similar quantum electrodynamics processes. 
These compound shifts test whether shower physics learned in one context can transfer to substantially different conditions.

The model autonomously adapts, guided solely by the objective function \cite{Amram:2024fjg}, from the compact \textsc{ILD} geometry to new geometric configurations such as the larger cylindrical configuration of \textsc{CaloChallenge} dataset 3, preserving fundamental particle shower physics while adjusting to changes in spatial scale and detector granularity.
For this study, electromagnetic shower physics is a good testing ground for geometry and scale adaptation since it is essentially particle-agnostic beyond the first interaction stage.

PEFT strategies are investigated to enhance sustainability by updating only parameter subsets. 
All training strategies are evaluated across varying downstream dataset sizes ($10^2$ to $10^5$ samples) to assess data efficiency when expensive \textsc{Geant4} simulation limits available training data.

\section{Datasets}
\label{sec:datasets}

\begin{figure}[htbp]
    \centering
    \includegraphics[width=0.99\linewidth]{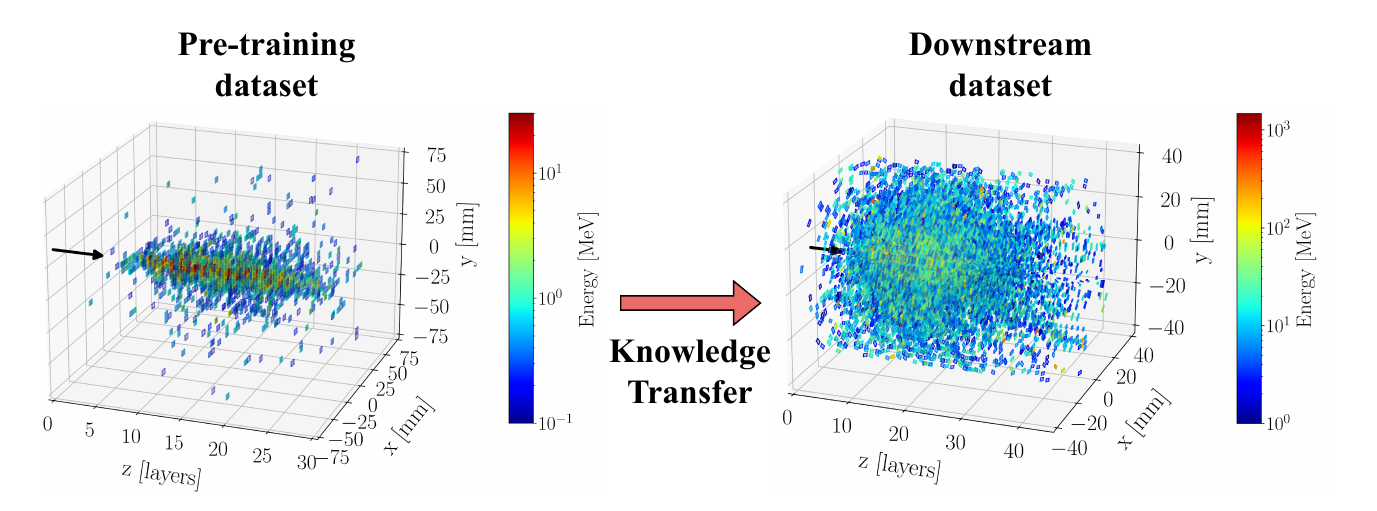}
    \caption{Representative electromagnetic shower event displays illustrating the domain shift. 
    Left: 81 GeV photon shower in the planar \textsc{ILD} detector. 
    Right: 913 GeV electron shower in the cylindrical \textsc{CaloChallenge} detector. 
    The cylindrical layer structure is visible in the curved distribution of energy deposits along the longitudinal axis. Data representation from Ref. \cite{Buss:2024orz}.}
    \label{fig:event_displays}
\end{figure}

This study employs two distinct electromagnetic shower datasets generated 
through \textsc{Geant4} simulations. 
The first dataset comprises photon showers simulated in the \textsc{ILD} detector, a realistic detector design developed for potential 
construction at the International Linear Collider for model pre-training, while the second contains electron showers in a cylindrical calorimeter geometry for transfer learning evaluation for downstream.
Figure \ref{fig:event_displays} shows visually the datasets considered in this study.

\subsection{Pre-training dataset}

This section describes the pre-training dataset used before task-specific fine-tuning. 
The approach employs the electromagnetic calorimeter (ECAL) datasets from  Ref. \cite{Buhmann:2023bwk}, utilising these pre-trained representations as the starting point. 
The pre-training dataset consists of $524$k\footnote{The pre-training dataset is available at \href{https://zenodo.org/records/10044175}{https://zenodo.org/records/10044175}.} photon showers with incident energy uniformly distributed between $10$ and $90$ GeV, simulated in the ILD~\cite{ILDConceptGroup:2020sfq}.
 
The ILD ECAL features $30$ layers alternating between tungsten absorbers ($2.1$ mm thick for the first $20$ layers, $4.2$ mm for the last $10$) and silicon sensors ($0.5$ mm thick with $5$ mm $\times$ $5$ mm readout cells).
Data representation employs two coordinate systems: a local system [$X$, $Y$, $Z$] centred at the photon's impact position, and a global ILD system [$X'$, $Y'$, $Z'$], with photons originating at [$X'=0$, $Y'=1811.3$ mm, $Z'=4$ mm] travelling along $Y'$.
The energy depositions from Geant4 (so called steps) are pre-clustered by layer and projected onto a grid with $36$ times higher resolution than the physical calorimeter ($0.83$ mm $\times$ $0.83$ mm cells), reducing approximately $20,000$ points per shower by a factor of roughly $7$.
Cluster positions are normalised to  [$-1$, $1$] within a bounding box from $-200$ mm to $200$ mm in $X$ and $Y$.

\subsection{Downstream dataset}
\label{sec:downstream}

For task-specific fine-tuning, this study employs \textsc{Dataset 3} \cite{calochallenge_dataset3} from the Fast Calorimeter Simulation Challenge (\textsc{CaloChallenge}) \cite{calochallenge2022}, designed to facilitate deep generative model development for calorimeter simulation \cite{Krause:2024avx}. \textsc{Dataset 3} contains electron showers with log-uniform incident energies from 1~GeV to 1~TeV, simulated using the geometry from the Par04 example of Geant4 \cite{geant4_par04}.

This geometry represents an idealised cylindrical calorimeter consisting of 90 concentric cylinders alternating between absorber material (1.4~mm of tungsten (W)) and active material (0.3~mm of silicon (Si)), contrasting with the planar ILD geometry. 
The calorimeter has an inner radius of 800~mm and a depth of 153~mm, with perpendicular showers positioned in the central $\eta = 0$ section. In the frame of reference considered in this study, each voxel along the $y$-axis corresponds to two physical layers (W-Si-W-Si) with a length of $\Delta z = 3.4$~mm (equivalent to $0.8 X_0$ of the absorber), resulting in $45$ readout layers compared to $30$ in the pre-training dataset. Showers are segmented into 18 radial and $50$ azimuthal bins, yielding $900$ voxels per layer and $40,500$ voxels per shower.
This segmentation, combined with the broader energy range, produces point clouds that can exceed three times the size of pre-training data at the highest energies. 

To enable effective transfer learning, the \textsc{CaloChallenge} dataset undergoes preprocessing to align with the pre-training format (detailed in Appendix~\ref{app:preprocessing}). Key steps include cylindrical smearing to convert voxelized deposits into continuous point clouds, sampling-fraction reversal to recover raw energy depositions, and point-based ordering for batch assembly efficiency.

The combination of geometric transformation from planar to cylindrical layout, energy distribution change from uniform (10--90 GeV) to log-uniform (1--1000 GeV), and differences in detector granularity creates a challenging transfer learning scenario that tests whether representations learned from ILD photon showers generalize to fundamentally different downstream conditions. The dataset is split into 100,000 samples for training with 10,000 samples reserved for validation and testing.

\section{Experiments}
\label{sec:results}

To assess the transfer learning capabilities for cross-geometry shower generation, this study examines how pre-trained representations influence downstream performance across different detector configurations. 
The experimental design isolates the contribution of learned physics knowledge by evaluating different training strategies and fine-tuning approaches across varying training dataset sizes. 
Random training examples are sampled from the full training set of \textsc{CaloChallenge}.

The training methodology is adapted according to computational requirements and model complexity. 
For the \textsc{PointWise} point cloud diffusion model generator \cite{Buhmann:2023kdg}, which represents the most computationally expensive component, all training strategies are evaluated to assess the trade-off between adaptation effectiveness and computational cost. 
Detailed training hyperparameter specifications are provided in  Appendix \ref{app:model hyperparms}.

For the \textsc{ShowerFlow} model, which determines the total number of points for point cloud post-diffusion calibration, only full fine-tuning is employed due to its relatively modest computational requirements during training. 
This approach leverages the complete learned representations while maintaining training efficiency for this less computationally complex architectural component.

\subsection{Evaluation Metrics}
\label{subsec:Evaluation metrics}

Generative models for calorimeter simulation must accurately reproduce statistical distributions of the training data. 
This evaluation employs hit-level and shower-level observables to assess model fidelity, comparing distributions between ground truth and generated samples using physically meaningful metrics, shown in Table \ref{tab:observables}.

Two complementary statistical metrics quantify agreement between generated samples and \textsc{Geant4} reference data :

\begin{description}
    \item[\textbf{Kullback-Leibler divergence}] provides a robust distributional comparison across the entire observable range: 
    
    \begin{equation}
    KL(P||Q) = \sum_{i} P_i \log \left(\frac{P_i}{Q_i}\right)
    \label{eq:quantile_kl_divergence}
    \end{equation}

    where $P_i$ and $Q_i$ represent the probabilities of reference and generated samples in the $i$-th bin.
    Bins are defined by reference distribution quantiles rather than fixed widths, ensuring uniform sensitivity across the observable range, and preventing dominance by high-density regions while capturing tail behaviour.
    The KL divergence is computed using \texttt{scipy.stats.entropy} \cite{Virtanen:2020scipy}.

    \item[\textbf{Wasserstein-1 distance}] offers a symmetric measure of distributional similarity based on optimal transport theory:
    \begin{equation}
    W_1(P, Q) = \min_{\pi \in \Pi(P,Q)} \sum_{i,j} |x_i - x_j| \pi(x_i, x_j)
    \label{eq:wasserstein_distance}
    \end{equation}
    Here, $\Pi(P, Q)$ denotes the joint coupling distributions with marginals $P$ and $Q$. 
    This metric quantifies the minimal cost of transforming one distribution into another, providing geometrically interpretable measures to small shifts from calorimeter resolution effects.
    The Wasserstein-1 distance is computed using \texttt{scipy.stats.wasserstein\_distance} \cite{Virtanen:2020scipy}.
\end{description}

\begin{table}[htbp]
\centering
\caption{Observables for evaluating generated calorimeter shower fidelity.}
\label{tab:observables}
\begin{tabular}{@{}lp{10cm}@{}}
\toprule
\textbf{Observable} & \textbf{Description} \\
\midrule

Voxel Energy Spectrum & Distribution of energy depositions across all voxels  \\[0.5em]
Energy ratio & Total measured energy summed over all voxels divided by incident energy\\[0.5em]
Visible Energy & Total energy deposition per shower  \\[0.5em]

Occupancy & Fraction of active voxels in a shower \\[0.5em]
Longitudinal Profile & Energy-weighted distribution along calorimeter layers \\[0.5em]
Radial Profile & Energy-weighted distribution of distances from the incident point \\[0.5em]
\bottomrule
\end{tabular}
\end{table}

These complementary metrics offer a comprehensive assessment of generation quality: quantile KL divergence provides uniform sensitivity across the full observable range, while Wasserstein distance measures overall distributional similarity.

\subsection{\textsc{ShowerFlow} Transfer Learning \& Post-Diffusion Calibration}
\label{sec:showerflow}

\begin{figure}[htbp]
    \centering
    \includegraphics[width=0.89\linewidth]{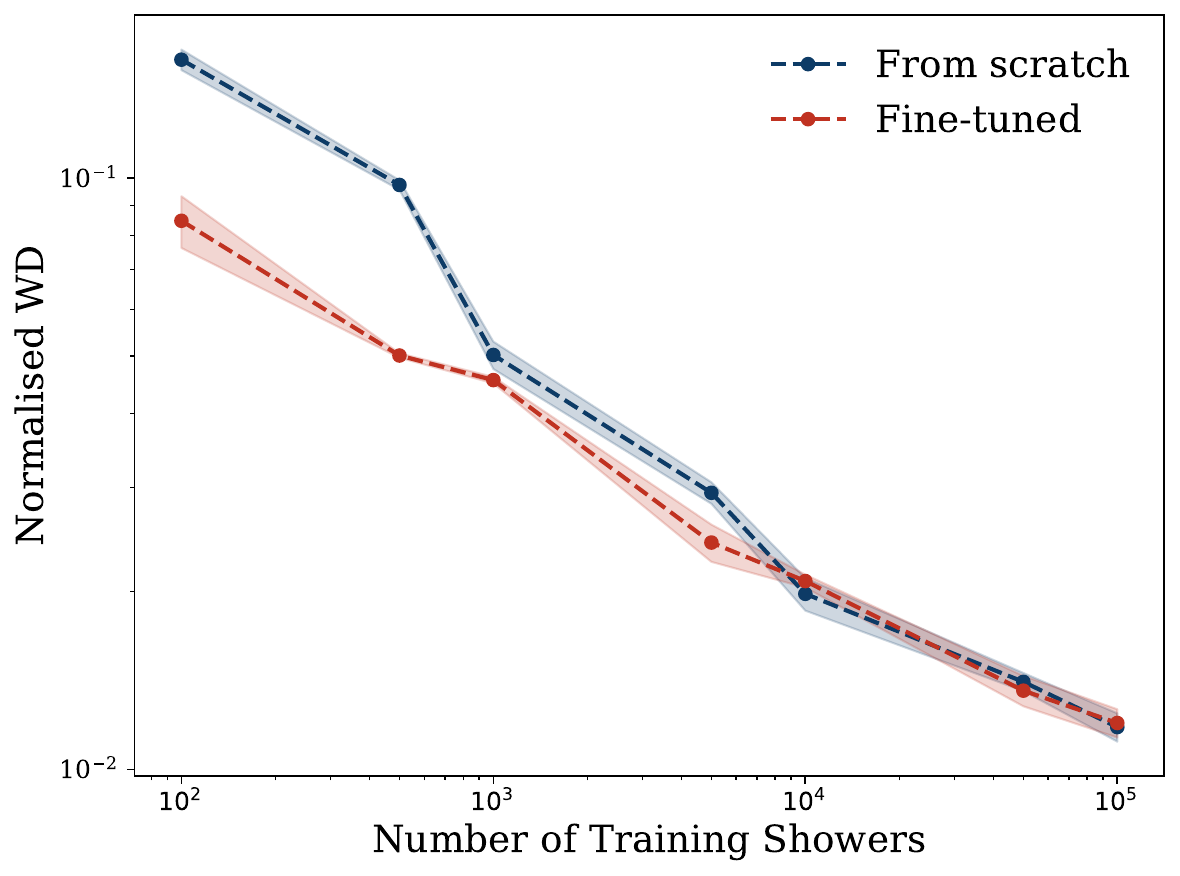}
    \caption{\textsc{ShowerFlow} transfer performance measured by normalised Wasserstein distance between generated and reference point-count distributions, averaged across all 45 calorimeter layers. Each point represents the median performance across five independent training runs with different random seeds. Error bands show the standard deviation across seeds. Evaluation is performed on the full 10,000-sample validation set. \textsc{Fine-Tuning} from ILD-pretrained weights substantially outperforms training \textsc{From Scratch} in low-data regimes.}
    \label{fig:shower_flow_wd_final_results1-1000}
\end{figure}

\textsc{ShowerFlow} predicts the point counts per layer $N_{z,i}$, i.e. the number of energy deposits in layer $i$, that condition \textsc{PointWise Net}'s point cloud generation. This model is trained exclusively for occupancy prediction and subsequent occupancy-based calibration, rather than energy per layer calibration as in Ref.~\cite{Buhmann:2023kdg}. 
To correct systematic biases in generated occupancy, we apply an energy-dependent calibration to the predicted point counts\footnote{This effect arises from information loss when projecting generated point clouds onto the detector's geometric configuration. To compensate, we oversample the number of points per layer using the polynomial calibration function.} that matches the relationship between total point count and occupancy fraction (active voxels) in generated versus reference showers. We fit cubic polynomials $p_{\text{data}}(O)$ and $p_{\text{gen}}(O)$ relating occupancy to point counts for reference and generated data respectively, then apply the transformation $N_{\text{cal}} = p_{\text{gen}}^{-1}(p_{\text{data}}(N_{\text{gen}}))$ to map generated counts through the reference occupancy relationship. 
Unlike the original manual approach, this automatically adapts to new datasets, with the calibrated counts $N_{\text{cal}}$ and counts per layer $N_{z,i,\text{cal}}$ subsequently conditioning the diffusion sampling.

The pretrained ILD model has 30 layers while \textsc{CaloChallenge} has 45 layers, creating a dimensional mismatch for the normalising flow architecture that cannot dynamically expand.
To bridge this gap, we model the additional 15 layers using log-normal distributions with parameters $(\mu, \sigma)$ estimated from 100 randomly sampled \textsc{CaloChallenge} showers, corresponding to the smallest dataset size we evaluate. During fine-tuning, the model predicts counts for the original 30 layers using pretrained weights, while the extra 15 layers are initialised from these log-normal distributions and then learned. \\
Formally, the total predicted count is $N_{\text{gen}} = \sum_{i=1}^{30} N_{z,i}^{\text{ILD}} + \sum_{i=31}^{45} N_{z,i}^{\text{adapted}}$, where the first term uses ILD pretrained backbone representations and the second term adapts to the new geometry.

Figure~\ref{fig:shower_flow_wd_final_results1-1000} shows that fine-tuning consistently outperforms training from scratch across all dataset sizes (see Appendix~\ref{app:showerflow} for detailed per layer histograms and convergence analysis, as well as the KL metric evaluation).
The benefit is clear in low-data regimes ($< 10^3$ samples) where pretrained representations provide essential inductive bias, reducing overfitting despite the architectural workaround for layer mismatch.

This layer extension strategy generalises to arbitrary detector configurations with different numbers of calorimeter layers.
For a target detector with $L_{\text{target}}$ layers and a pre-trained model with $L_{\text{pre}}$ layers, the approach requires modifications only in the \textsc{ShowerFlow} component: when $L_{\text{target}} > L_{\text{pre}}$, additional layers are initialised from parametric distributions estimated from a small target domain sample, as demonstrated here.
When $L_{\text{target}} < L_{\text{pre}}$, the excess output dimensions can be discarded or merged. 
The \textsc{PointWise Net} diffusion model requires no architectural changes, as it operates on individual points conditioned on layer indices rather than fixed-dimensional outputs, making it inherently flexible to varying layer counts.

\subsection{Cross-Calorimeter Performance}

All results in this section employ the complete generation pipeline: \textsc{ShowerFlow} predicts point counts $N_{z,i}$ per layer, which then condition \textsc{PointWise Net}'s diffusion-based point cloud generation. 
 For the comparison between \textsc{from scratch} and \textsc{full fine-tuned} models (subsection \ref{subsec:scratch_vs_ft}), both \textsc{ShowerFlow} and \textsc{PointWise Net} are trained with the same strategy. For parameter-efficient methods (subsection  \ref{sec:peft}), \textsc{ShowerFlow} is always fully fine-tuned due to its modest computational cost, while \textsc{PointWise Net} employs various PEFT techniques. 

Adapting a pre-trained model to a new detector geometry requires careful consideration of inference time constraints. 
Since the target application requires fast and scalable point cloud generation, we focus exclusively on fine-tuning techniques that preserve the original inference speeds; methods such as adapters \cite{pmlr-v97-houlsby19a} are then excluded. Only methods that retain the original inference graph are considered: partial fine-tuning, \textsc{BitFit}, and Low-Rank Adaptation (\textsc{LoRA}).
This constraint ensures practical deployment in latency-critical applications while demonstrating that \textsc{LoRA} and \textsc{BitFit} extend effectively beyond language models to point cloud diffusion tasks.

All performance metrics represent Wasserstein distances computed for six physics observables (see Section \ref{subsec:Evaluation metrics}).
We aggregate these using the geometric mean to ensure balanced evaluation across observables with different scales:

\begin{equation}
    \bar{y}_{jk} = \left( \prod_{i=1}^6 y_{ijk} \right)^{1/6},
    \label{eq:geom_mean}
\end{equation}

\begin{figure}[htbp]
    \centering
    \begin{subfigure}[b]{0.94\linewidth}
        \centering
        \includegraphics[width=\linewidth]{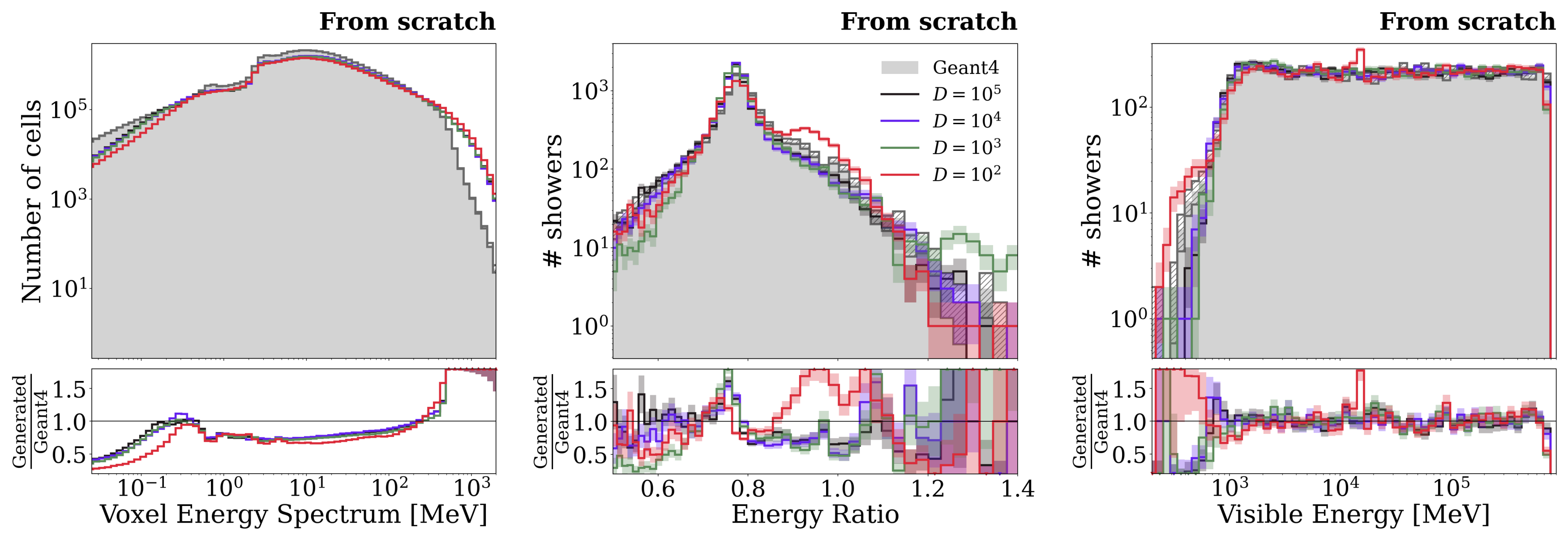}
    \end{subfigure}
    \vspace{0.2cm}
    \begin{subfigure}[b]{0.94\linewidth}
        \centering
        \includegraphics[width=\linewidth]{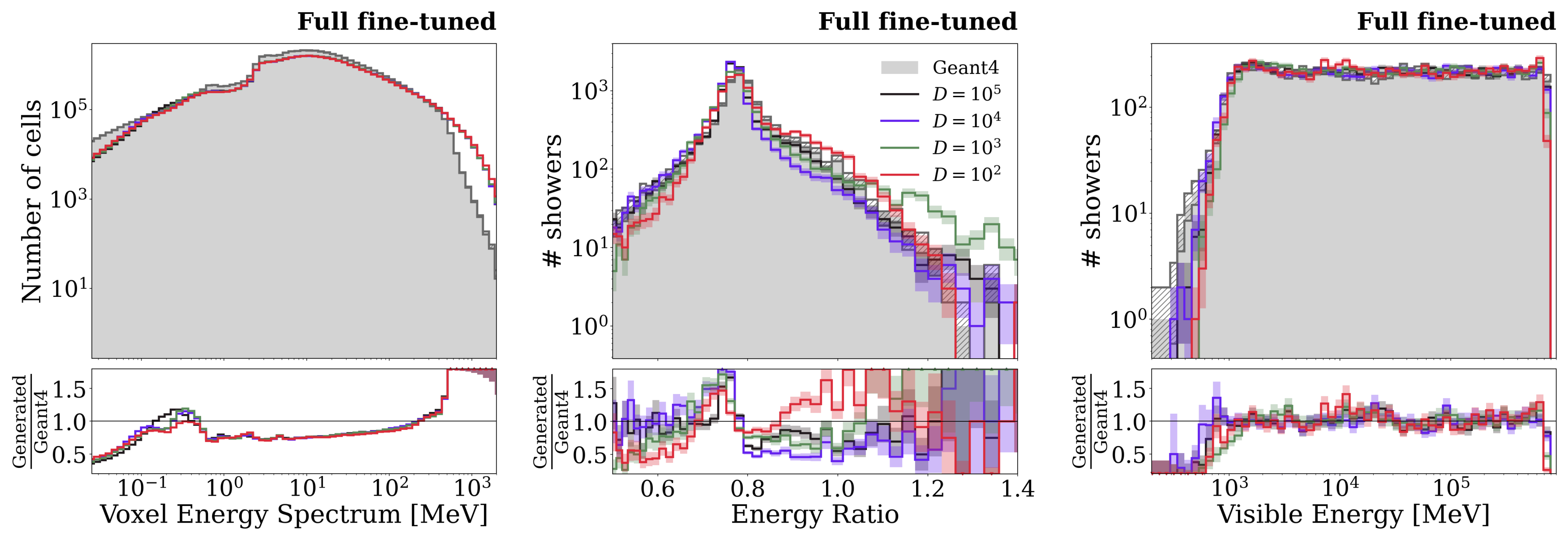}
        
        \caption{Distributions: cell energy spectrum (left), total deposited energy over incident energy (centre), visible energy (right).}
    \end{subfigure}

    \begin{subfigure}[b]{0.94\linewidth}
        \centering
        \includegraphics[width=\linewidth]{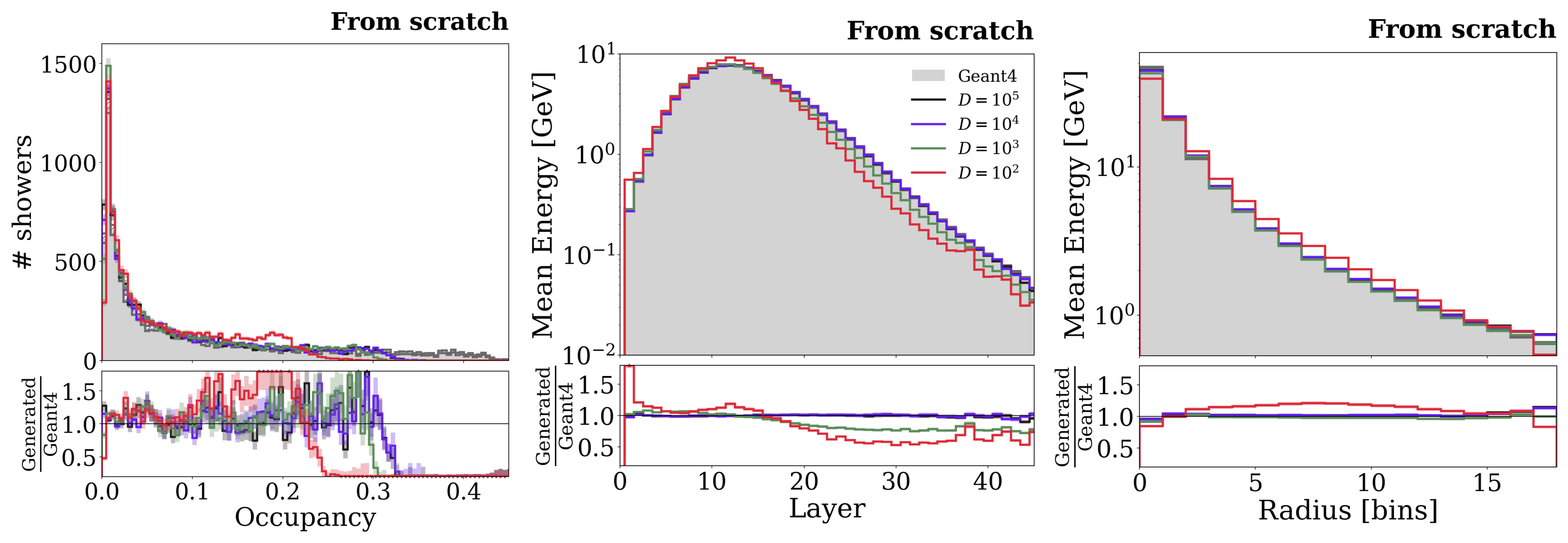}
    \end{subfigure}
    \vspace{0.2cm}
    \begin{subfigure}[b]{0.94\linewidth}
        \centering
        \includegraphics[width=\linewidth]{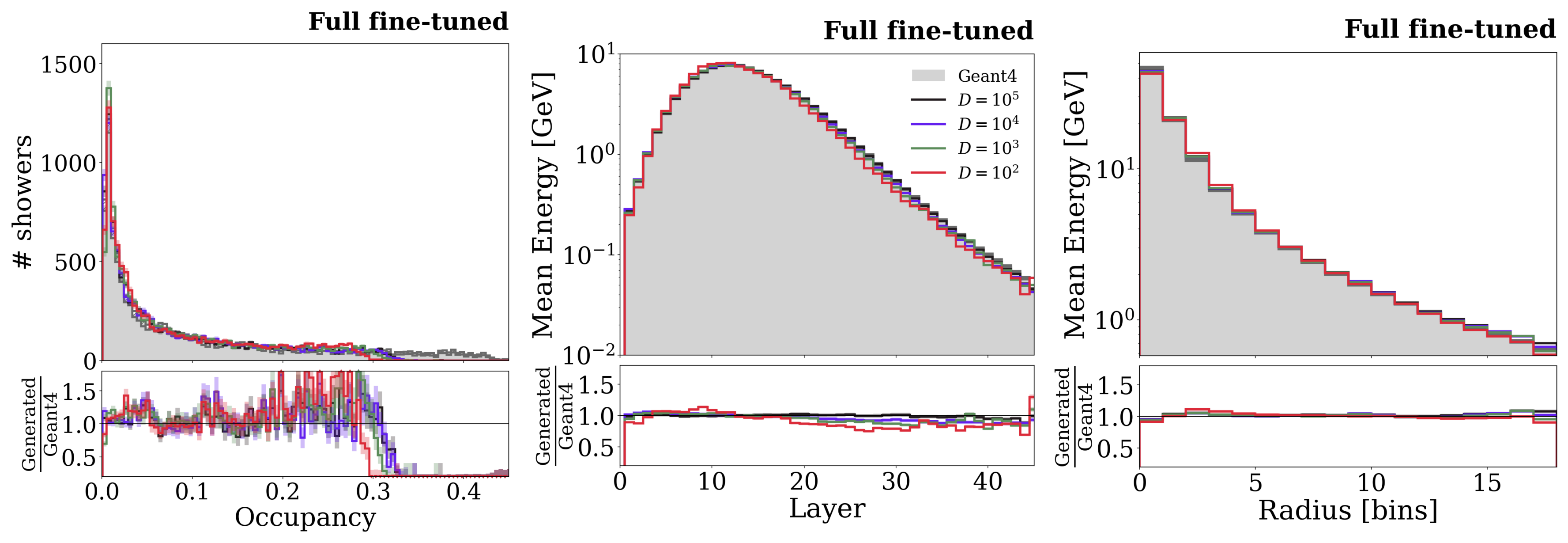}

        \caption{Distributions: occupancy (left), longitudinal (centre), radial profile (right).}
    \end{subfigure}

    \caption{\textsc{Geant4} vs generated showers at training sizes $D$. Top rows: \textsc{from scratch}; bottom rows: \textsc{full fine-tuned}. All histograms from $10,000$ events with energy logarithmically distributed from $1-1000$ GeV. Bottom panels show \textsc{Geant4} ratios. The error band corresponds to the statistical uncertainty in each bin.}
    \label{fig:comparison_scratch_finetuned}
\end{figure}

where each training method $j$ across the six physical observables $i$ is calculated for different training shower sizes $k$. 
This prevents any single metric from dominating the evaluation while maintaining sensitivity to performance variations. While this aggregation provides useful guidance and quantitative benchmarks, we emphasise examining individual observables directly, as aggregated metrics can obscure important physics specific performance patterns. 
The geometric mean serves primarily to guide overall assessment, while detailed observable analysis reveals the true model behaviour. Further consideration on Equation \ref{eq:geom_mean} and its error propagation is detailed in Appendix \ref{app:geom_mean}.

\subsubsection{From scratch vs Full fine-tuning}
\label{subsec:scratch_vs_ft}

We now compare the two training strategies introduced in Section \ref{sec:tl_framework}. In this comparison, both \textsc{ShowerFlow} and \textsc{PointWise Net} are either trained \textsc{from scratch} with random initialisation, or \textsc{full fine-tuned} from ILD pretrained weights.
We use the term full fine-tuning to distinguish this from the parameter-efficient methods examined in Section \ref{sec:peft}.

\begin{figure}[htbp]
    \centering
    \includegraphics[width=0.99\linewidth]{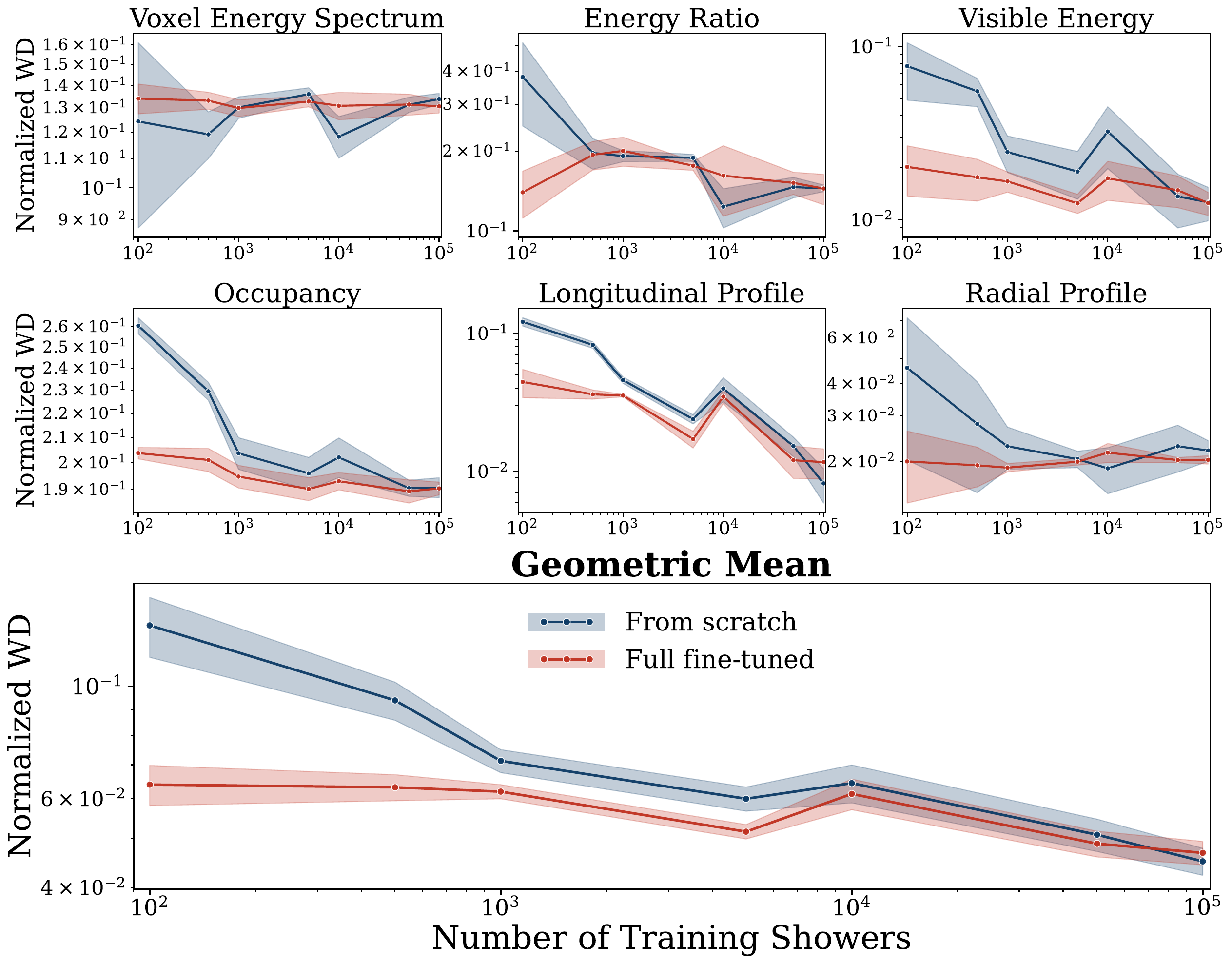}
    \caption{Wasserstein evaluation metrics for shower generation for the six physical observables across different dataset sizes. The resulting bands represent averages over five independent seeds with RMS uncertainty bands. A residual at $10^4$ samples remains visible in some individual observables.}
    \label{fig:scratch_FT_final_results_wd_1-1000Gev}
\end{figure}

Figure \ref{fig:comparison_scratch_finetuned} shows distinct performance patterns across training dataset sizes. 
The voxel energy spectrum reveals minimal differences between training strategies and dataset sizes, with both approaches yielding similar distributions, regardless of whether pre-training is used. 
In addition, both approaches generate excessively high-energy voxel deposits (>$100$ MeV) compared to \textsc{Geant4}, likely due to the point cloud projection occasionally concentrating multiple hits into a single voxel.
Despite this limitation, the observable appears to be learned effectively even without transfer learning, suggesting that the point cloud representation naturally captures the energy deposition patterns independent of the source detector.
This contrasts with geometric observables, like longitudinal and radial profiles, where pre-training provides clear advantages.
Occupancy is underestimated at high values due to information loss during the projection from point clouds to regular cell geometry. 
The \textsc{Full fine-tuned} training shows superior performance in longitudinal and radial profiles, particularly at low data regimes, demonstrating better adaptation of shower structure to the new geometry. 
Additionally, the improved visible energy performance in the \textsc{fine-tuned} model indicates more stable energy ratio modelling, especially crucial when training data is limited.

Figure~\ref{fig:scratch_FT_final_results_wd_1-1000Gev} quantifies the transfer learning advantage. 
With only $10^2$ training samples, \textsc{full fine-tuned} model achieves a geometric mean Wasserstein distance of $0.064 \pm 0.006$ compared to $0.132 \pm 0.018$ for \textsc{from scratch} training.
Despite the large variance in the baseline, the $\sim51\%$ reduction in mean WD demonstrates statistically significant transfer learning benefits in data constrained scenarios. 
This benefit diminishes with the increase in training data.

Individual observables show differential sensitivity to transfer learning. 
Longitudinal and radial profiles benefit most, as geometric features learned from \textsc{ILD} transfer effectively despite detector differences.
The voxel energy spectrum shows minimal improvement, likely because point clouds inherently provide dense sampling for this observable regardless of training set size.

At $10^4$ training samples, a residual increase in Wasserstein distance remains visible in some individual observables, most notably the longitudinal profile and visible energy. 
After retraining with non-overlapping data partitions to ensure fully independent runs, this effect persists, indicating a genuine model behaviour rather than a statistical artefact.
This effect is confined to individual metrics and does not impact the aggregate geometric mean.
The overall trend continues to demonstrate a clear transfer learning advantage in the low-data regime ($<10^3$ samples).

\subsubsection{Parameter-Efficient Fine-Tuning Strategies}
\label{sec:peft}

Beyond full fine-tuning, we evaluate adaptation methods that update only a subset of parameters in \textsc{PointWise Net} while preserving the original inference architecture. 
These techniques may offer crucial advantages for multi-detector deployment and computational efficiency.
As pretrained models scale and become more general-purpose, the computational cost of retraining all parameters for each detector configuration becomes increasingly impractical, particularly when considering deployment across multiple experimental setups. 
The study presented in this section is the first application of PEFT methods to a pretrained model in the context of fast particle shower simulations.

\begin{description}
    \item[\textbf{\textsc{BitFit}}]\cite{ben-zaken-etal-2022-bitfit} represents the most parameter-efficient approach, training only bias terms while freezing all weights. This method modifies $17\%$ of the model parameters by recalibrating activation thresholds throughout the network, thereby adjusting response patterns for the target detector geometry.

    \item[\textbf{\textsc{Top2}}] fine-tuning freezes the feature extraction layers and updates only the final two layers\footnote{\textit{Final layers} in this context refer to those closer to the output.} as well as the time-step layer. This approach tests whether the earlier layers contain reusable representations that enable accurate generation with minimal adaptation. This configuration was selected through systematic ablation studies that examined various combinations of layers, revealing that updating the top two and the time-embedding layers provides the optimal balance between expressivity and efficiency.

    \item[\textbf{\textsc{LoRA}}]\cite{hu2021loralowrankadaptationlarge} introduces low-rank decomposition matrices that adapt pretrained representations through additive updates. We employ rank $106$, selected based on the comprehensive analysis in Appendix~\ref{app:lora}, where we demonstrate that \textsc{CaloChallenge} requires higher ranks than typical NLP applications due to the high-dimensional nature of particle shower transformations.
\end{description}

\begin{figure}[htbp]
    \centering
    \includegraphics[width=0.99\linewidth]{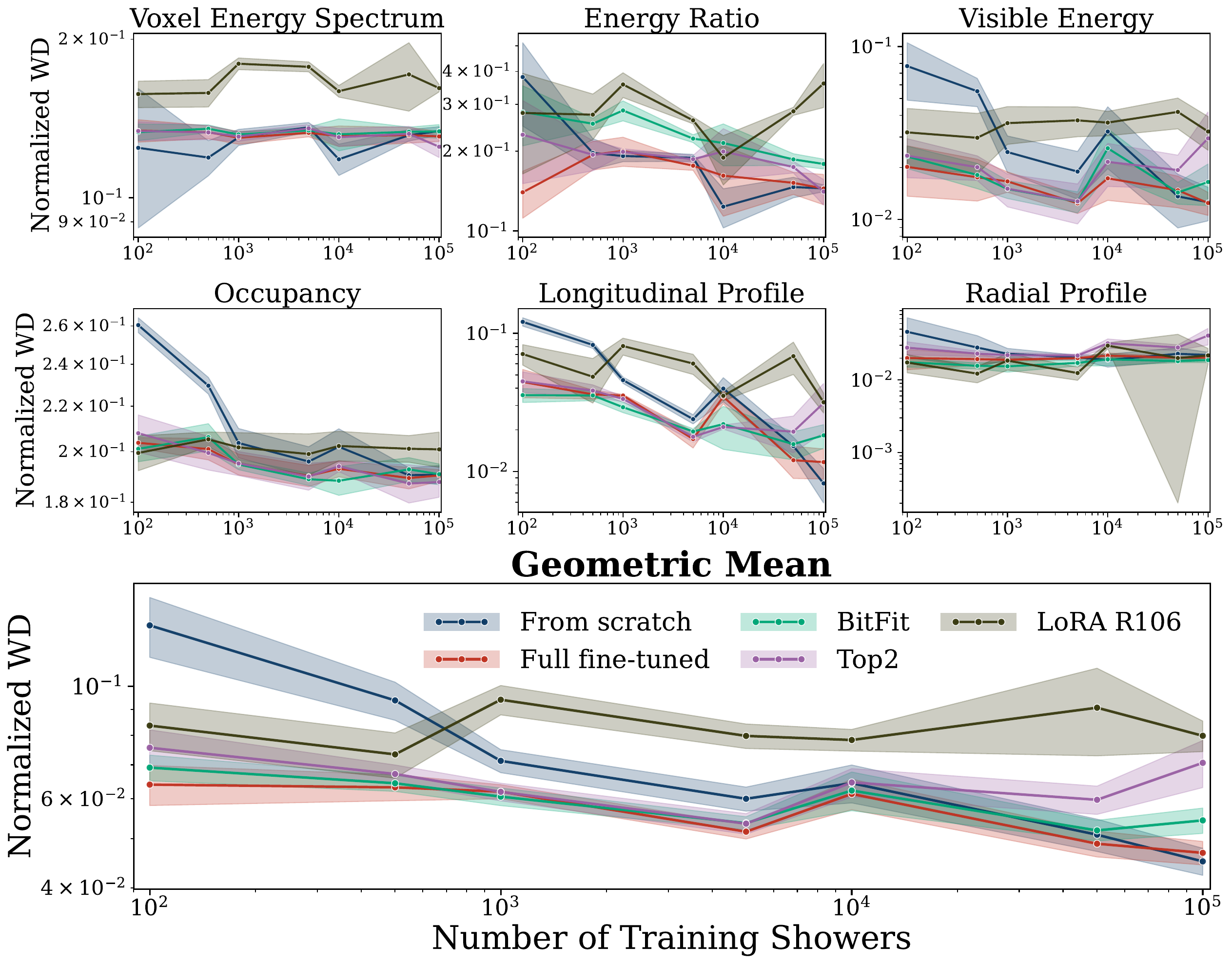}
    \caption{Parameter-efficient fine-tuning performance across training data volumes. Wasserstein distances evaluated on generated showers. Uncertainties represent standard error across three independent training runs with different random seeds.}
    \label{fig:PEFT_final_results_wd_1-1000Gev}
\end{figure}

Table~\ref{tab:performance} presents quantitative comparisons across methods and dataset sizes. \textsc{BitFit} achieves performance within $5\%$ of full fine-tuning on the geometric mean Wasserstein distance while updating only $17\%$ of trainable parameters.
\textsc{Top2} fine-tuning with $44\%$ of parameters shows comparable results, suggesting that adaptation primarily occurs in higher layers while lower layers remain largely transferable.
\textsc{LoRA} exhibits degraded performance despite utilising $52\%$ of parameters, with consistently poor results at intermediate data scales and particular degradation in the voxel energy spectrum, though showing comparable performance for energy ratio and occupancy observables.

\begin{table}
\centering
\caption{Performance comparison across training strategies. Geometric mean Wasserstein distance values $\times10^{-2}$ for readability. Uncertainties show standard error over five seeds. At $D = 10^4$, training sets are constructed from non-overlapping data partitions to ensure independence across runs.}
\label{tab:performance}
\small
\begin{tabular}{l|c|cccc|c}
\toprule
\multirow{2}{*}{\textbf{Method}} & \multirow{2}{*}{\textbf{Params (\%)}} & \multicolumn{4}{c|}{\textbf{Training Dataset Size}} & \multirow{2}{*}{\textbf{Mean}} \\
 & & 10$^2$ & 10$^3$ & 10$^4$ & 10$^5$ \\
\midrule
\textsc{From scratch} & 100\% & 13.2±1.8 & 7.1±0.4 & 6.4±0.6 & \textbf{4.5±0.3} & 7.8±0.5 \\
\textbf{\textsc{Full fine-tuned}} & 100\% & \textbf{6.4±0.6} & 6.2±0.2 & \textbf{6.1±0.4} & 4.7±0.3 & \textbf{5.9±0.2} \\ \midrule
\textsc{BitFit} & 17\% & 6.9±0.4 & \textbf{6.1±0.2} & 6.2±0.5 & 5.4±0.3 & 6.2±0.2 \\
\textsc{Top2} & 44\% & 7.6±0.6 & 6.2±0.2 & 6.5±0.4 & 7.1±0.8 & 6.8±0.3 \\
\textsc{LoRA R106} & 52\% & 8.4±0.9 & 9.4±0.6 & 7.8±0.4 & 8.0±0.6 & 8.4±0.3 \\
\bottomrule
\end{tabular}
\end{table}

The results reveal several important patterns.
At small data scales ($10^2$), pre-training provides clear benefits across all methods, with transfer learning reducing Wasserstein distance by $51\%$ compared to training \textsc{from scratch}. The intermediate data regime ($10^3$--$10^4$) shows more complex behaviour, with minor variations in relative performance that may reflect sampling effects and the interplay between pre-training bias and target domain adaptation. At the largest scale ($10^5$), the performance gap narrows as sufficient data allows even \textsc{from scratch} training to converge effectively.

\textsc{LoRA}'s consistent underperformance warrants specific discussion. Unlike its success in NLP tasks, \textsc{LoRA} struggles with calorimeter simulation even at rank 106. Our analysis in Appendix~\ref{app:lora_svd} reveals that weight updates in shower physics exhibit high intrinsic dimensionality across network layers, with some requiring ranks exceeding 200 for accurate reconstruction. This fundamental mismatch between \textsc{LoRA}'s low-rank assumption and the complexity of physics transformations explains its limited effectiveness.

The success of \textsc{BitFit} and \textsc{Top2} methods suggests that effective adaptation for \textsc{CaloChallenge} operates through two mechanisms: recalibrating activation patterns via bias adjustments and refining high-level feature combinations in final layers. Both approaches preserve the learned representations while allowing targeted modifications for detector-specific characteristics.

\begin{table}[htbp]
\centering
\caption{Computational cost for each training method, measured on a single NVIDIA A100 80GB PCIe GPU 
with batch size 64. Per-step timings averaged over 3 epochs after 1 warmup epoch. 
Average GPU power draw is approximately constant across configurations ($\sim$278\,W).}
\label{tab:computational_cost}
\small
\begin{tabular}{l|c|c|c|c}
\toprule
\textbf{Method} & \textbf{Params (\%)} & \textbf{ms/step} & \textbf{Peak VRAM (GB)} & \textbf{Conv. steps ($\times 10^3$)$^\ddagger$} \\
\midrule
\textsc{From scratch} & 100\% & 62.3 $\pm$ 0.3 & 16.4 & $\sim$500 \\
\textsc{Full fine-tuning} & 100\% & 62.3 $\pm$ 0.3 & 16.4 & $\sim$\textbf{100} \\
\textsc{BitFit} & 17\% & \textbf{49.0 $\pm$ 0.1} & \textbf{12.6} & $\sim$200 \\
\textsc{Top2} & 44\% & 54.5 $\pm$ 0.4 & 14.8 & $\sim$250 \\
\textsc{LoRA R106} & 52\%$^\dagger$ & 86.9 $\pm$ 0.4 & 13.7 & $\sim$50 \\
\bottomrule
\end{tabular}
\vspace{1mm}

\footnotesize{$^\dagger$ Relative to original model parameters (504,920). \textsc{LoRA} introduces additional parameters, bringing the total to 777,128.}\\
\footnotesize{$^\ddagger$ Approximate convergence steps, defined as the first step within 5\% of each method's best geometric mean Wasserstein distance. Values vary across dataset sizes, metrics, and seeds; reported here as indicative estimates.}
\end{table}

Table~\ref{tab:computational_cost} quantifies the computational cost of each training strategy. \textsc{BitFit} reduces per-step training time by 21\% and peak memory by 23\% compared to full fine-tuning, while \textsc{Top2} achieves intermediate reductions of 12\% and 10\% respectively. \textsc{LoRA} incurs a 40\% increase in training time despite fewer trainable parameters, due to the additional low-rank matrix operations. Full fine-tuning converges in approximately 100k steps compared to 500k for training from scratch, making pre-training beneficial both in final performance and in total training cost. Combining per-step cost with convergence, \textsc{BitFit} offers the most efficient total training time ($\sim$200k $\times$ 49\,ms $\approx$ 2.7\,h) compared to full fine-tuning ($\sim$100k $\times$ 62\,ms $\approx$ 1.7\,h) and from scratch ($\sim$500k $\times$ 62\,ms $\approx$ 8.6\,h), while using 23\% less peak memory. Since GPU power draw is approximately constant across all configurations, per-step energy consumption scales proportionally with training time.

These findings have relevant implications for deploying generative models across diverse detector configurations. 
The reduced memory requirements of parameter-efficient methods may enable multi-geometry adaptation without proportional increases in storage. 
By freezing most parameters, these techniques accelerate convergence and mitigate catastrophic forgetting~\cite{MichaelMcCloskeyforgetting}, essential properties for continual learning across evolving detector designs. 
As calorimeter models scale up and become more general, our results indicate that successful adaptation strategies might respect the high-dimensional nature of physics data, favouring threshold recalibration and selective layer updates over aggressive low-rank compression. 
These empirical findings challenge the universal applicability of low-rank adaptation methods and motivate the development of physics-aware parameter-efficient techniques.

\section{Conclusions and Outlook}
\label{sec:conclusions}

This study explores single-detector pre-training on point cloud representations as a path for generalisable cross-geometry transfer learning in calorimeter simulation. 
Our findings demonstrate that meaningful transfer learning is achievable even from single-geometry pre-training.

The main findings are that, in low-data regimes ($10^2$ samples), pre-training on ILD photon showers enables adaptation to the \textsc{CaloChallenge} electron shower task, yielding a statistically significant $51\%$ performance improvement over training \textsc{from scratch}.
Among parameter-efficient methods, \textsc{BitFit} achieves performance within $5\%$ of full fine-tuning using only $17\%$ of parameters, while \textsc{LoRA} shows limited effectiveness even at rank $106$. 
Our post-hoc singular value analysis provides theoretical insight into why \textsc{LoRA} struggles, suggesting that particle shower transformations may have higher intrinsic dimensionality than typical NLP tasks.

Several limitations constrain our conclusions. 
A residual increase in Wasserstein distance at $D=10^4$ persists in some individual observables across independent runs, suggesting a genuine model behaviour at this intermediate data scale rather than a statistical artefact. 
This effect does not impact the aggregate geometric mean.
Computational benchmarking demonstrates that parameter efficiency translates into practical cost reductions: \textsc{BitFit} reduces per-step training time by $21\%$ and peak memory by $23\%$, while full fine-tuning converges in approximately 100k steps compared to 500k for training from scratch.
Most importantly, without direct comparison to multi-detector pre-training approaches, we cannot claim relative performance against existing foundation model approaches.

Despite these limitations, this work contributes to understanding transfer learning in calorimeter simulations.
The success of \textsc{BitFit} and selective layer fine-tuning suggests that adaptation primarily involves targeted recalibration rather than fundamental representation changes.
In scenarios where multi-detector datasets are unavailable or computational resources are limited, single-detector pre-training offers an adequate starting point for rapid prototyping.

Future work should pursue several directions. 
First, developing more generalizable pre-training strategies that leverage point cloud representations across different calorimeter geometries and energy ranges would strengthen the foundation model approach.
Second, systematic comparisons with multi-detector pre-training approaches would establish relative performance benchmarks. 
Finally, extending this framework to hadronic showers and mixed particle types would test the generalizability of transfer learning in more complex scenarios.

\section*{Code Availability}

The code for this study can be found under \href{https://github.com/FLC-QU-hep/CaloTransfer}{https://github.com/FLC-QU-hep/CaloTransfer}.

\acknowledgments

We thank Katja Krüger for valuable comments on the manuscript.
This research was supported in part by the Maxwell computational resources operated at Deutsches
Elektronen-Synchrotron DESY, Hamburg, Germany. 
This project has received funding from the
European Union’s Horizon 2020 Research and Innovation programme under Grant Agreement No
101004761. We acknowledge support by the Deutsche Forschungsgemeinschaft under Germany’s
Excellence Strategy – EXC 2121 Quantum Universe – 390833306 and via the KISS consortium
(05D23GU4, 13D22CH5) funded by the German Federal Ministry of Research, Technology, and
Space (BMFTR) in the ErUM-Data action plan.

\newpage

\appendix
\newpage
\section{Pre-processing}
\label{app:preprocessing}
To enhance transfer learning from the pre-trained model, the \textsc{CaloChallenge} dataset is aligned with the pre-training format via three preprocessing steps:

\begin{description}
    \item[Cylindrical smearing:] Voxelized energy depositions are converted into point clouds. 
    Each energy deposit, originally localised at the voxel centre, is spatially redistributed by sampling uniformly within the cylindrical boundaries of its host voxel.
    This process applies Gaussian noise to the radial ($r$) and azimuthal ($\phi$) coordinates while preserving the longitudinal ($z$) position, ensuring energy conservation within individual detector cells. 
    The smearing maintains the detector's cylindrical geometry while generating continuous spatial distributions that facilitate the training of diffusion models. 
    Figure~\ref{fig:cylindrical_smearing} illustrates this transformation for a representative shower with incident energy of $500.3$ GeV, showing the transition from discrete voxelized deposits to spatially smeared point clouds in the transverse plane of the calorimeter.
    
    \begin{figure}[htbp]
        \centering
        \includegraphics[width=.99\linewidth]{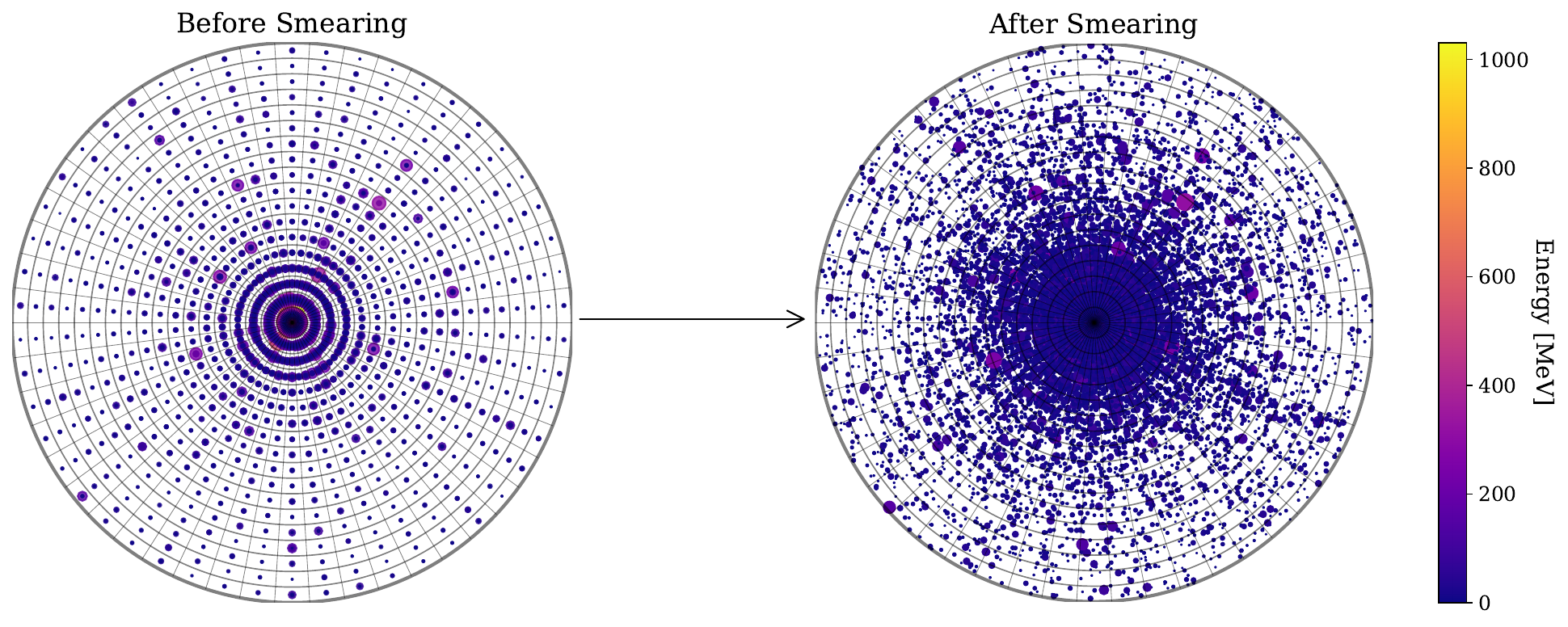}
        \caption{Cylindrical smearing transformation applied to electromagnetic shower data from \textsc{CaloChallenge}. The left panel shows the original voxelized energy depositions concentrated at voxel centres. 
        The right panel displays the result after cylindrical smearing, where energy deposits are spatially redistributed within their respective voxel boundaries using Gaussian noise in cylindrical coordinates. 
        The colour scale represents energy deposition values, and the concentric circles indicate the detector's cylindrical segmentation.}
        \label{fig:cylindrical_smearing}
    \end{figure}
    
    \item[Sampling-fraction reversal:] The normalisation applied to account for sampling fractions is inverted to recover raw energy depositions in the active material, matching the silicon-layer energy scale used during pre-training.
    
    \item[Point-based ordering:] Showers are sorted by point count to assemble mini-batches of similar complexity, replicating the efficiency gains observed in the original pre-training data version.
\end{description}

These procedures maintain the physical integrity of \textsc{CaloChallenge} showers while standardizing geometry, energy scale, and batch complexity, thereby facilitating effective cross-architecture knowledge transfer.

\section{Hyperparameters used in experiments }
\label{app:model hyperparms}

\begin{table*}[htbp]
\centering
\caption{\textsc{PointWise Net} settings and sampling parameters used across all training methods.}

\label{tab:general_hyperparameters}
\small
\begin{tabularx}{0.75\textwidth}{lX}
\toprule
\textbf{Category} & \textbf{Configuration} \\
\midrule
\multirow{5}{*}{\textbf{Training Setup}} 
& Batch Size: 64 \\
& Optimizer: RAdam \\
& LR Schedule: Linear (100K warmup → 300K decay) \\
& Maximum Gradient Steps: 1.1M \\
& Weight Decay: 0.01 \\
& Device: NVIDIA\textsuperscript{\textregistered} A100 \\

\midrule
\multirow{4}{*}{\textbf{EDM Configuration}} 
& KL Weight ($\beta$): $10^{-3}$ \\
& KLD Min: 1.0 \\
& Noise Schedule: Quadratic \\
& EMA: Inverse (power=0.6667, max=0.9999) \\
\midrule
\multirow{6}{*}{\textbf{Sampling}} 
& $\sigma_{\text{data}}$: 0.5 \\
& $\sigma$ Distribution: LogNormal($\mu=-1.2$, $\sigma=1.2$) \\
& ODE Solver: Heun \\
& Sampling Steps: 32 \\
& $\sigma_{\min}$ / $\sigma_{\max}$: 0.002 / 80.0 \\
& $\rho$ / $s_{\text{churn}}$ / $s_{\text{noise}}$: 7.0 / 0.0 / 1.0 \\
\bottomrule
\end{tabularx}
\end{table*}

\begin{table*}[htbp]
\centering
\caption{\textsc{PointWise Net} Learning rate schedules and method-specific parameters adapted to different dataset sizes.}
\label{tab:method_hyperparameters}
\small
\begin{tabularx}{0.9\textwidth}{ll|cccc}
\toprule
\multirow{2}{*}{\textbf{Method}} & \multirow{2}{*}{\textbf{Parameter}}
& \multicolumn{4}{c}{\textbf{Training Dataset Size}} \\
& & 10$^2$ & 10$^3$ & 10$^4$ & 10$^5$ \\
\midrule
\multirow{2}{*}{\textsc{From scratch}} 
& LR Start / End & \multicolumn{4}{c}{2e-4 / 1e-4} \\
& \# Gradient Steps & 250,000 & 1,000,000 & 500,000 & 750,000 \\
\midrule
\multirow{2}{*}{\textsc{Full Fine-tuned}} 
& LR Start / End & 5e-4/5e-5 & 1e-4/1e-5 & 2.5e-5/2.5e-6 & 5e-6/5e-7 \\
& \# Gradient Steps & 100,000 & 50,000 & 100,000 & 250,000 \\
\midrule
\multirow{2}{*}{\textsc{Top2 Fine-tuned}} 
& LR Start / End & 5e-4/5e-5 & 1e-4/1e-5 & 2.5e-5/2.5e-6 & 5e-6/5e-7 \\
& \# Gradient Steps & 1,000,000 & 500,000 & 750,000 & 750,000 \\
\midrule
\multirow{2}{*}{\textsc{BitFit}} 
& LR Start / End & 2e-3/2e-4 & 4e-4/4e-5 & 1e-4/1e-5 & 2e-5/2e-6 \\
& \# Gradient Steps & 1,000,000 & 750,000 & 500,000 & 500,000 \\
\midrule
\multirow{3}{*}{\textsc{LoRA R8}} 
& LR Start / End & 1e-3/1e-4 & 2e-4/2e-5 & 5e-5/5e-6 & 1e-5/1e-6 \\
& \# Gradient Steps & 250,000 & 10,000 & 100,000 & 200,000 \\
& LoRA $\alpha$ / $r$ & \multicolumn{4}{c}{8 / 8} \\
\midrule
\multirow{3}{*}{\textsc{LoRA R106}} 
& LR Start / End & 1e-3/1e-4 & 2e-4/2e-5 & 5e-5/5e-6 & 1e-5/1e-6 \\
& \# Gradient Steps & 100,000 & 100,000 & 10,000 & 50,000 \\
& LoRA $\alpha$ / $r$ & \multicolumn{4}{c}{106 / 106} \\
\bottomrule
\end{tabularx}
\end{table*}

Table \ref{tab:general_hyperparameters} shows the baseline configuration we used across all experiments, while Table \ref{tab:method_hyperparameters} details how we adapted learning rates for different training methods and dataset sizes. We report the median performance over 5 random seeds, with results taken from the best-performing epoch for each run. To maintain training stability while optimizing memory usage, we also implemented adaptive batch sizing following the approach of Keskar et al. \cite{Keskar:2016tzj}.

\begin{table}[htbp]
\centering
\caption{\textsc{ShowerFlow} model architecture and training configuration with dataset-dependent batch sizing.}
\label{tab:showerflow_hyperparameters}
\begin{tabularx}{0.7\textwidth}{lXc}
\toprule
\textbf{Category} & \textbf{Hyperparameter} & \textbf{ShowerFlow} \\
\midrule
\multirow{3}{*}{\textsc{Data}} 
& Pin Memory & True \\
& Workers & 4 \\
& Shuffle & True \\

\midrule
\multirow{6}{*}{\textsc{Architecture}} 
& Num Blocks & 2 \\
& Num Inputs & 45 \\
& Conditioning Inputs & 1 (Energy) \\
& Coupling Hidden Dims & [920, 920] \\
& Spline Hidden Dims & [368, 368] \\
& Spline Bins & 8 \\

\midrule
\multirow{8}{*}{\textsc{Training}} 
& Device & NVIDIA\textsuperscript{\textregistered} V100 \\
& Optimizer & Adam \\
& Scheduler & None \\
& Learning Rate & $1 \times 10^{-4}$ \\
& Batch Size$^{\dagger}$ & [64, 2048] \\
& Maximum Epochs & $1000$ \\
& Gradient Clipping$^{*}$ & $10^4$ → $5 \times 10^5$ \\
\bottomrule

\end{tabularx}
\vspace{2mm}\\
{\footnotesize $^{\dagger}$Batch size varies by training size:  $64$ ($10^2$ samples), $128$ ($10^3$ samples), $512$ ($10^4$ samples), $2048$ ($10^5$ samples).
$^{*}$Gradient clipping applied only for Fine-tuned, linearly increasing from $10^4$ to $5 \times 10^5$ over first $50$ epochs.}
\end{table}

Table \ref{tab:showerflow_hyperparameters} presents the configuration for the ShowerFlow model, which uses a different architecture and thus required its own optimization strategy. The batch sizes were scaled with dataset size to balance training efficiency and stability.

\section{\textsc{ShowerFlow} Transfer}
\label{app:showerflow}

\begin{figure}
    \centering
    \includegraphics[width=1\linewidth]{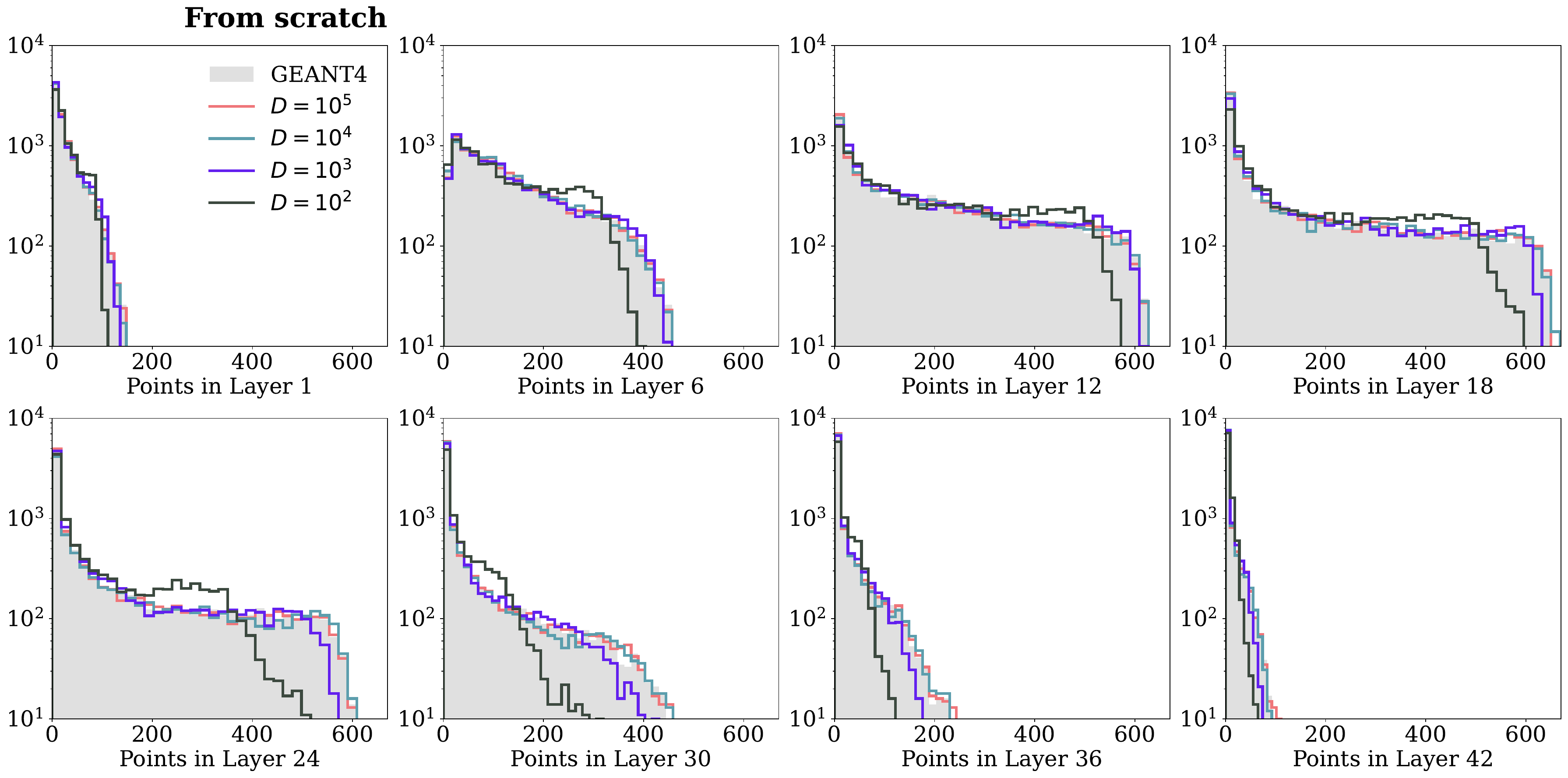}
    \caption{Histograms of points per layer for \textsc{CaloChallenge}: Geant4 reference (gray) versus \textsc{ShowerFlow} trained \textsc{from scratch} with varying dataset sizes. All distributions computed from $10,000$ showers with logarithmic energy sampling between $1$ and $1000$ GeV.}
    \label{fig:points_distr_scratch}
\end{figure}

\begin{figure}
    \centering
    \includegraphics[width=1\linewidth]{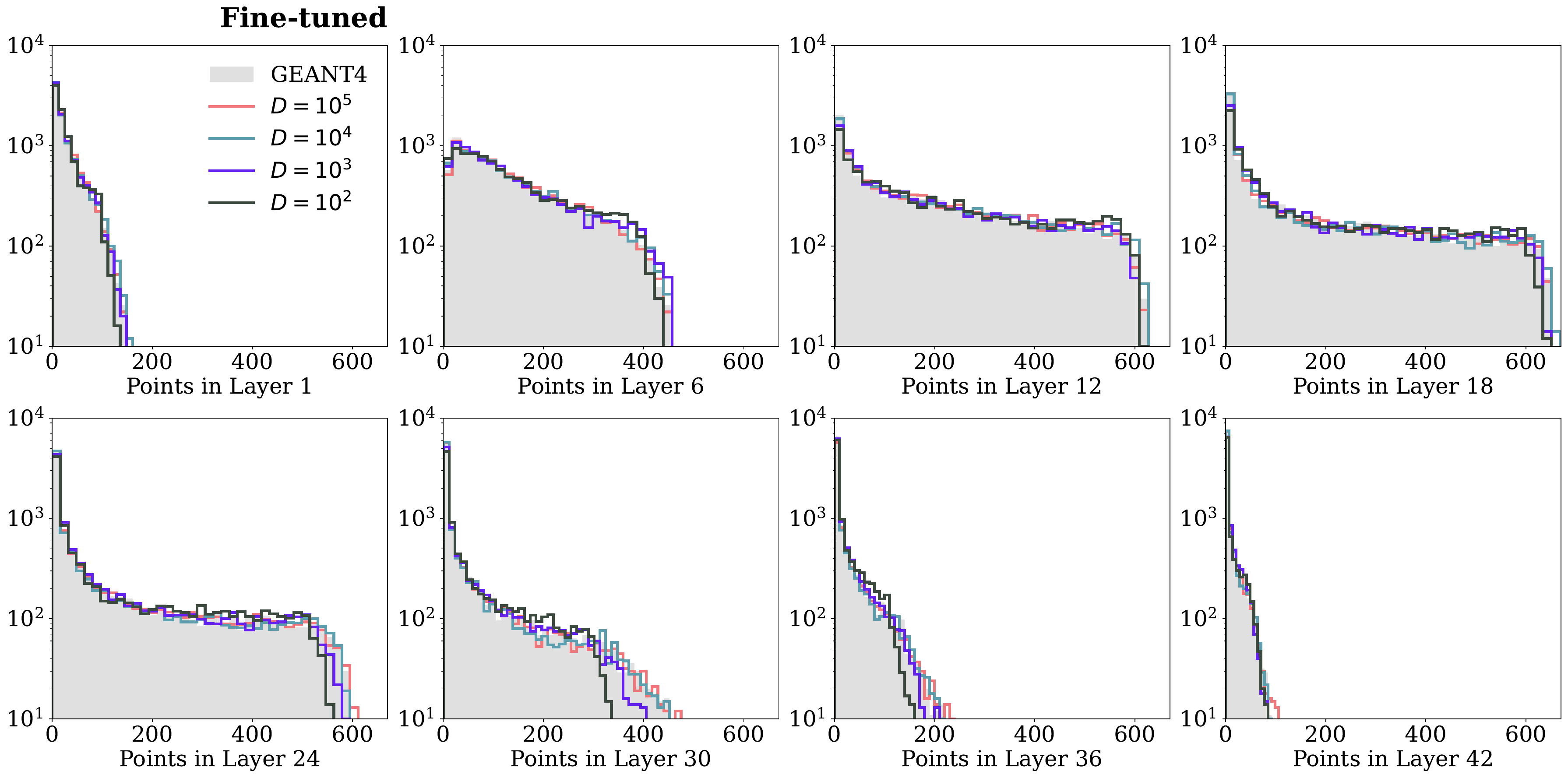}
    \caption{Histograms of points per layer for \textsc{CaloChallenge}: Geant4 reference (gray) versus \textsc{ShowerFlow} \textsc{finetuned} from ILD pre-training with varying dataset sizes. All distributions computed from $10,000$ showers with logarithmic energy sampling between $1$ and $1000$ GeV.}
    \label{fig:points_distr_finetruned}
\end{figure}

Figures \ref{fig:points_distr_scratch} and \ref{fig:points_distr_finetruned} show the distribution of points per layer generated by \textsc{ShowerFlow}. The shower development peaks between layers $10$ and $25$, where the electromagnetic cascade is most active, and the highest number of voxels are triggered.
Accurate modeling of these distributions is crucial since the points per layer serve as conditioning input for generating the full EM showers and calibrating the shower structure.
The \textsc{fine-tuned} model clearly outperforms the \textsc{from scratch} version in low data regimes, demonstrating successful knowledge transfer from the pre-training phase. 
This advantage becomes less pronounced as training data increases, since sufficient data allows the model to learn the distributions directly.

\begin{figure}[htbp]
    \centering
    \includegraphics[width=.95\linewidth]{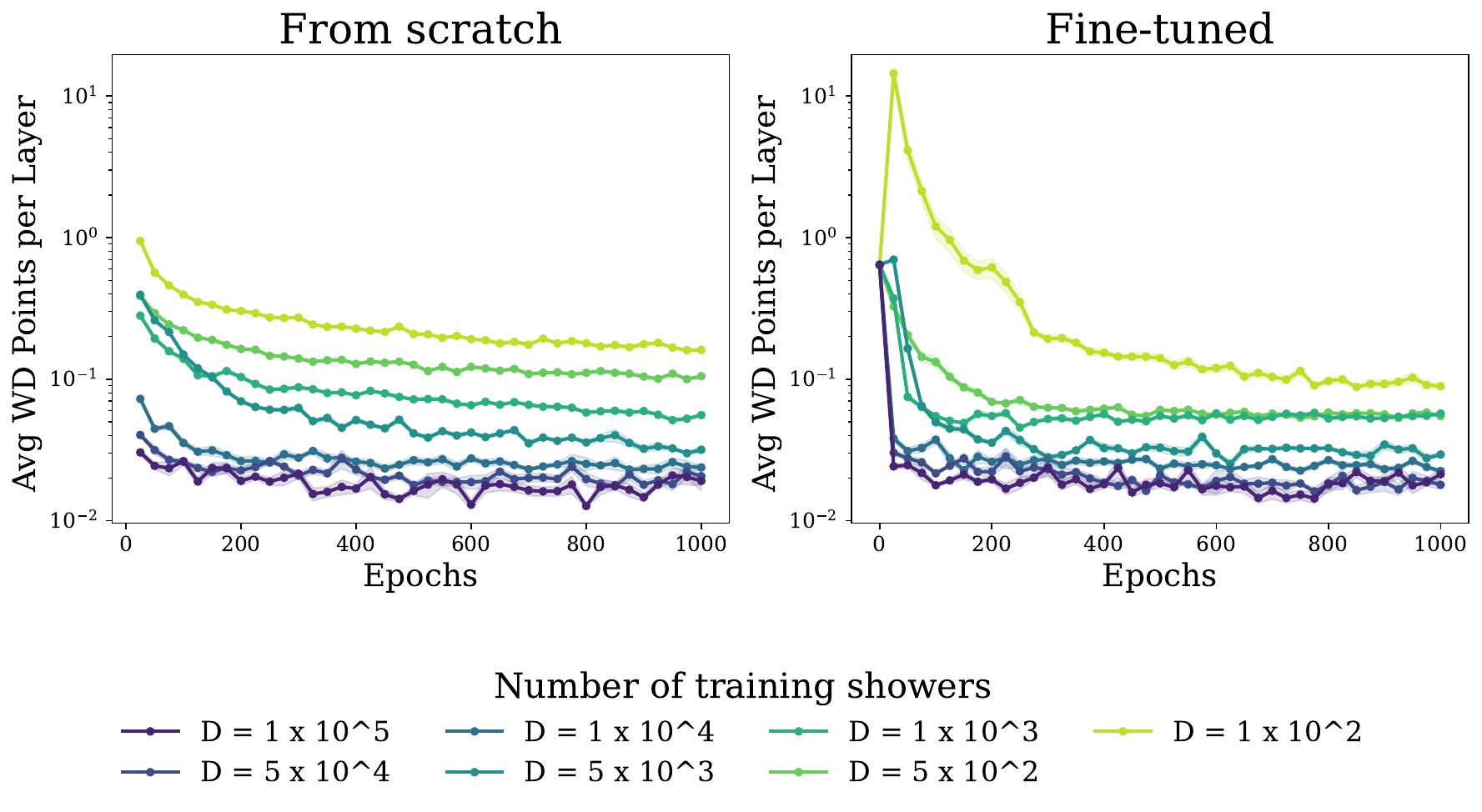}
    \caption{\textsc{ShowerFlow} convergence curves using Wasserstein distance. Each configuration averaged over $5$ random seeds. Epoch $0$ represents pretrained weights (\textsc{finetuned} version only).}
    \label{fig:sf_wd_convergence_curve_1-1000}
\end{figure}

\begin{figure}[htbp]
    \centering
    \includegraphics[width=.95\linewidth]{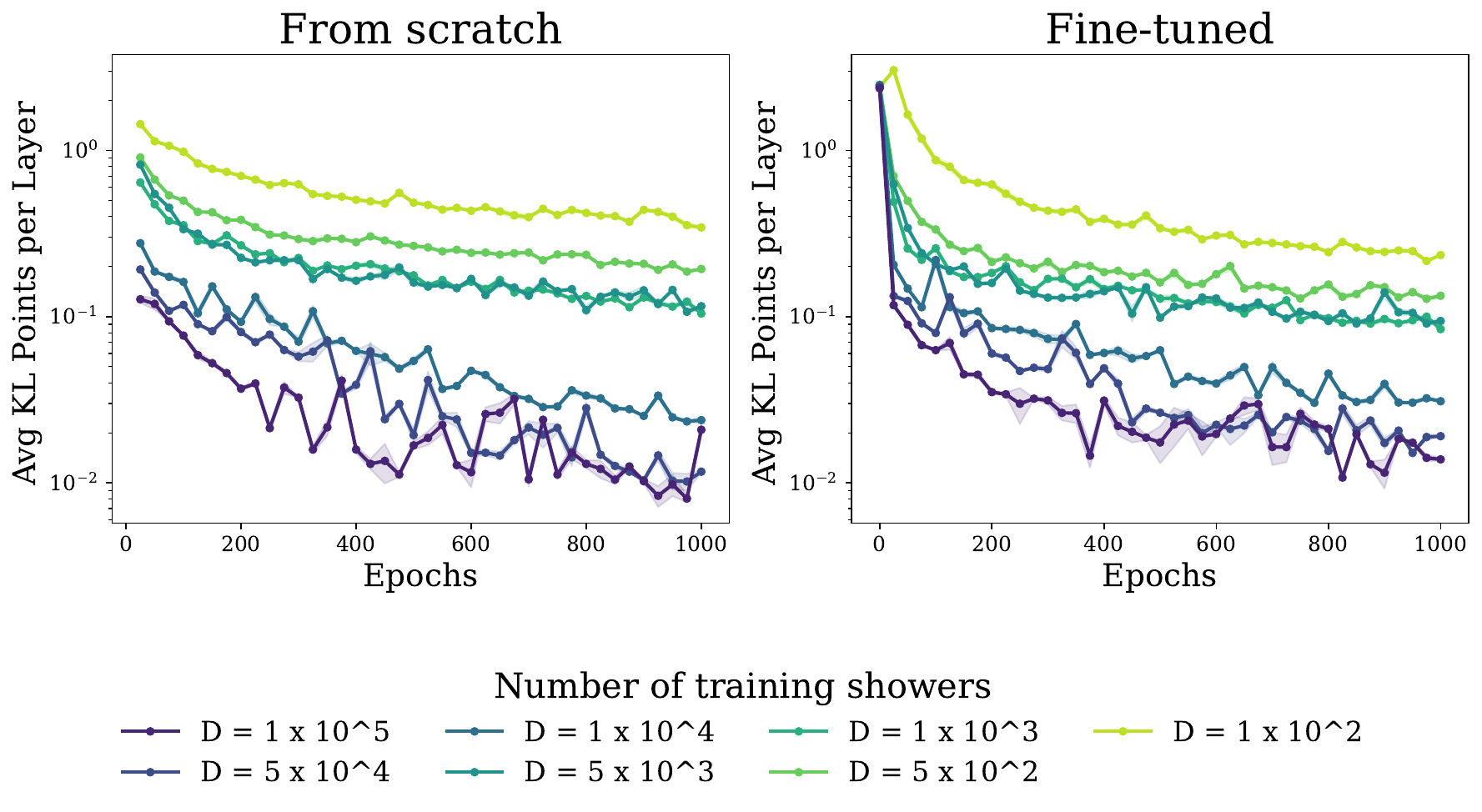}
    \caption{\textsc{ShowerFlow} convergence curves using KL divergence. Each configuration averaged over $5$ random seeds. Epoch $0$ represents pretrained weights (\textsc{finetuned} version only).}
    \label{fig:sf_kl_convergence_curve_1-1000}
\end{figure}

To select the optimal epoch for each configuration, we tracked the Wasserstein distance and KL divergence across training, as shown in Figures \ref{fig:sf_wd_convergence_curve_1-1000} and \ref{fig:sf_kl_convergence_curve_1-1000}. 
Given the training instability and overfitting risk in low data regimes, we selected epochs based on the minimum averaged metric across validation samples rather than single point validation loss minima.

\begin{figure}[htbp]
    \centering
    \includegraphics[width=0.89\linewidth]{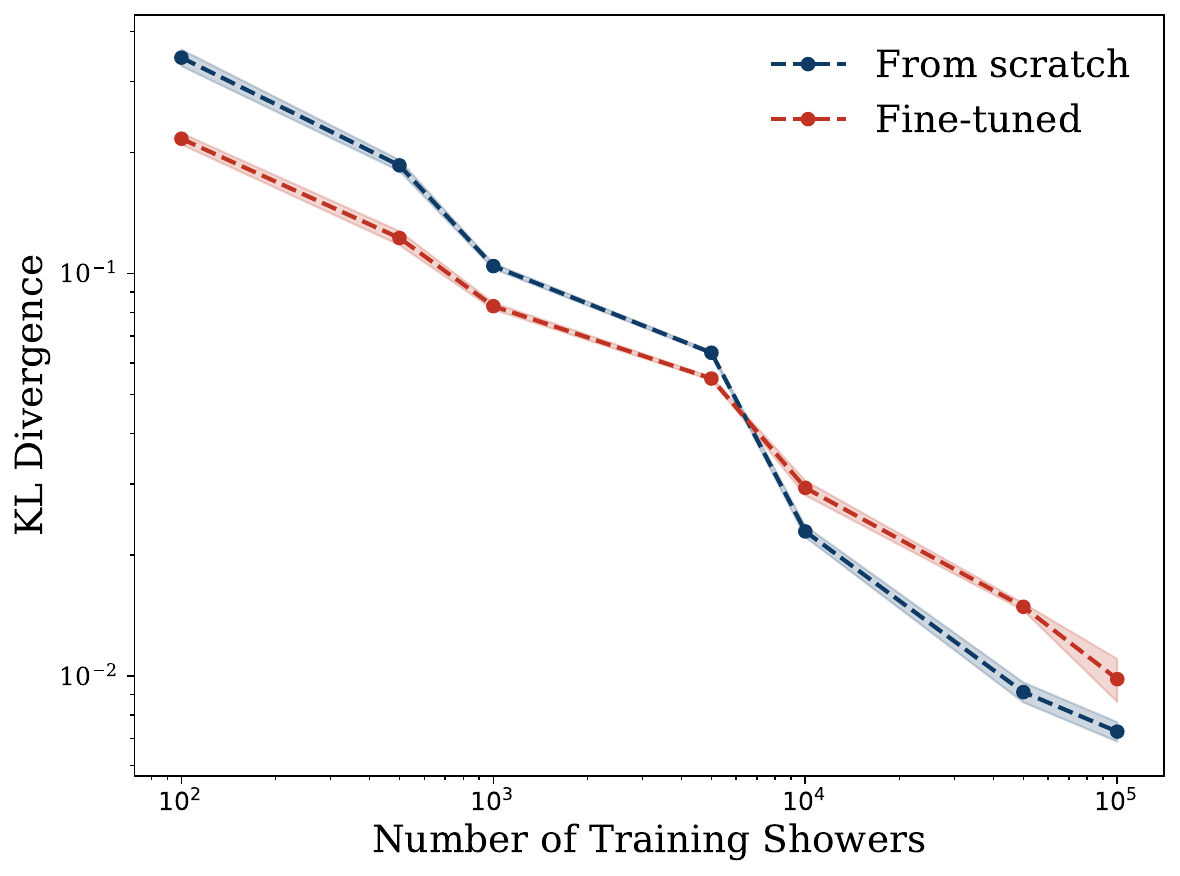}
    \caption{\textsc{ShowerFlow} transfer learning performance measured by KL divergence averaged across all calorimeter layers. Fine-tuning significantly outperforms training \textsc{from scratch} in low data regimes. Results averaged over five random seeds.}
    \label{fig:shower_flow_kl_final_results1-1000}
\end{figure}

Figure \ref{fig:shower_flow_kl_final_results1-1000} presents the final performance when epochs are selected using KL divergence, while Figure \ref{fig:shower_flow_wd_final_results1-1000} shows selection based on WD. 
Both metrics reveal that transfer learning provides substantial benefits primarily in low data regimes ($<5 \times 10^3$ samples), with comparable or slightly reduced performance at larger dataset sizes, where the model has sufficient data to learn from scratch.

\section{Further plots}
\label{app:complete model}

This section of the appendix presents comprehensive evaluation metrics that complement the main results.
Section \ref{app:peft_histos} provides detailed histogram comparisons for all parameter-efficient fine-tuning methods, while Section \ref{app:kl_evaluation} presents Kullback-Leibler divergence analysis as an alternative metric to validate the Wasserstein distance findings.

\subsection{PEFT histograms}
\label{app:peft_histos}

\begin{figure}[htbp]
    \centering
    \begin{subfigure}[b]{0.94\linewidth}
        \centering
        \includegraphics[width=\linewidth]{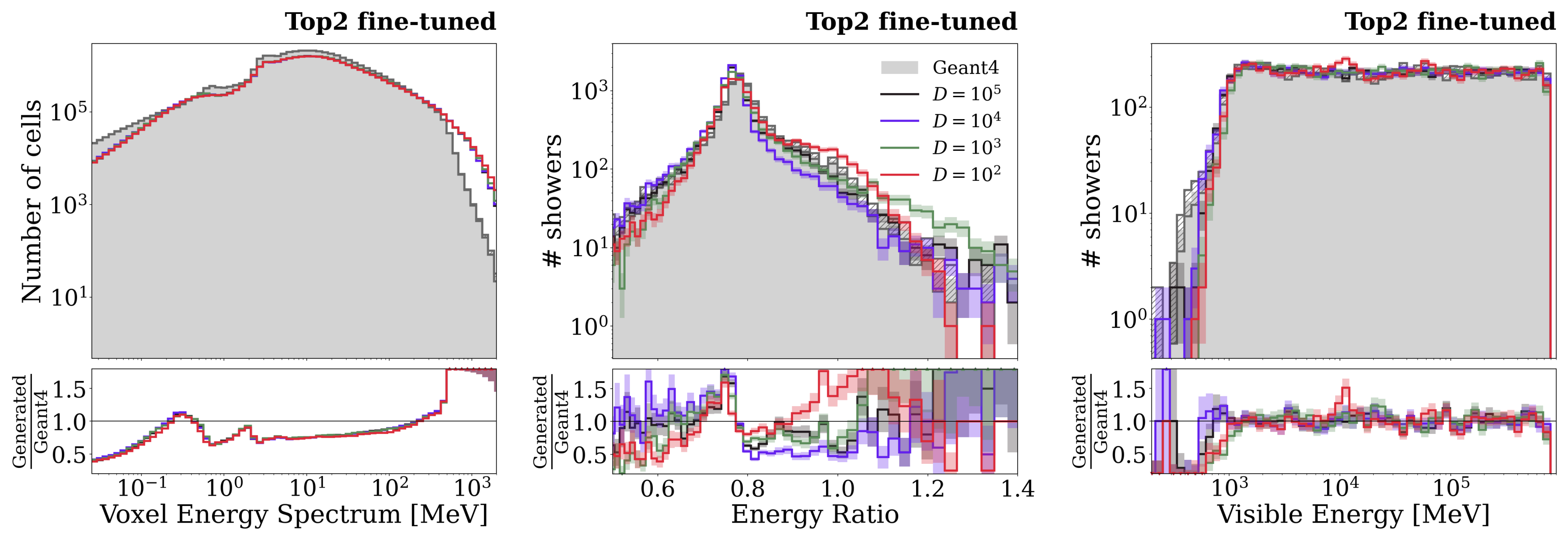}
    \end{subfigure}
    \vspace{0.2cm}
    \begin{subfigure}[b]{0.94\linewidth}
        \centering
        \includegraphics[width=\linewidth]{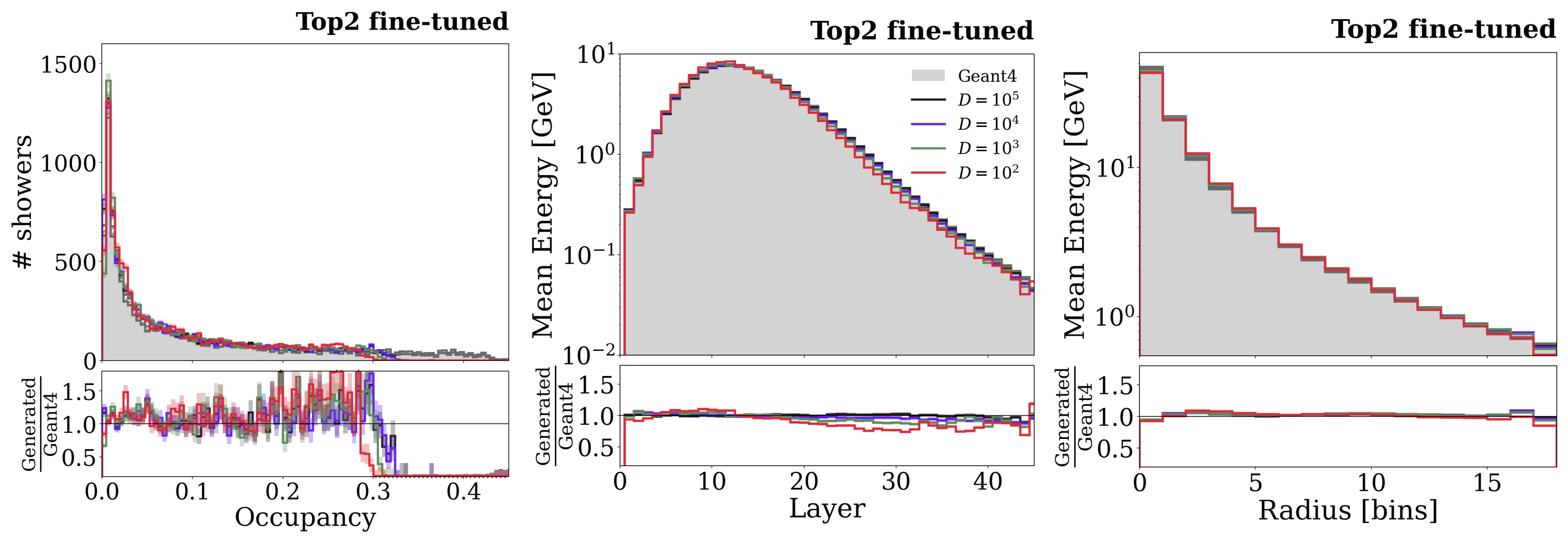}
    \end{subfigure}
    \vspace{0.2cm}
    \begin{subfigure}[b]{0.94\linewidth}
        \centering
        \includegraphics[width=\linewidth]{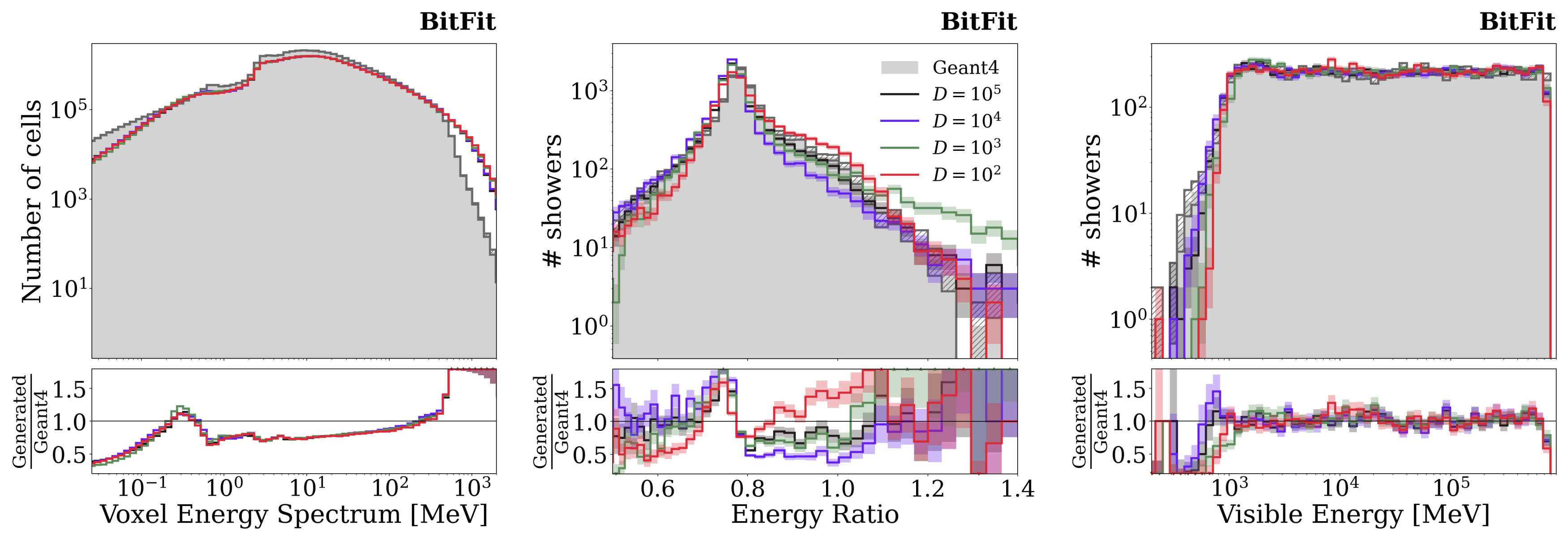}
    \end{subfigure}
    \vspace{0.2cm}
    \begin{subfigure}[b]{0.94\linewidth}
        \centering
        \includegraphics[width=\linewidth]{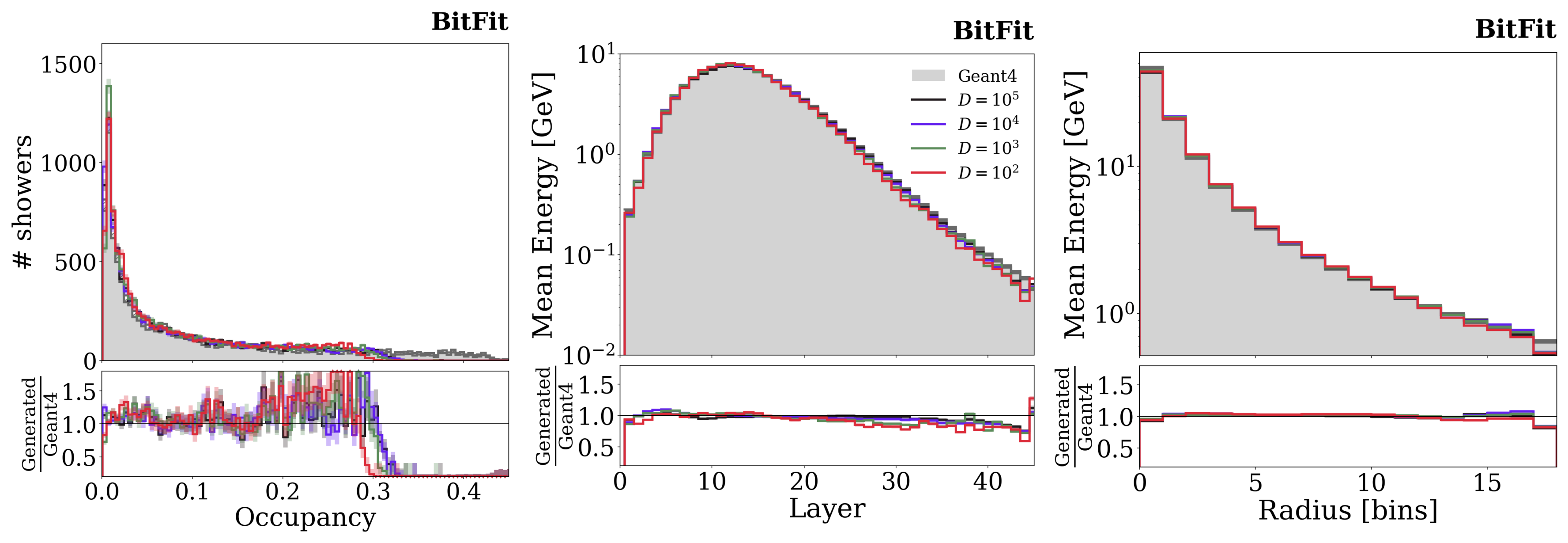}
    \end{subfigure}
    \caption{\textsc{Geant4} vs generated showers for parameter-efficient methods at training sizes $D$. 
    A comprehensive distribution analysis of generated showers for \textsc{Top2 fine-tuned} (first two rows), and \textsc{BitFit} (last two rows).
    All histograms from 10,000 events with energies logarithmically distributed from 1 to 1000 GeV. 
    Bottom panels show \textsc{Geant4} ratios with statistical uncertainties.
    The error band represents the statistical uncertainty in each bin.}
    \label{fig:comparison_peft_methods}
\end{figure}

\begin{figure}[htbp]
    \centering
    \begin{subfigure}[b]{0.94\linewidth}
        \centering
        \includegraphics[width=\linewidth]{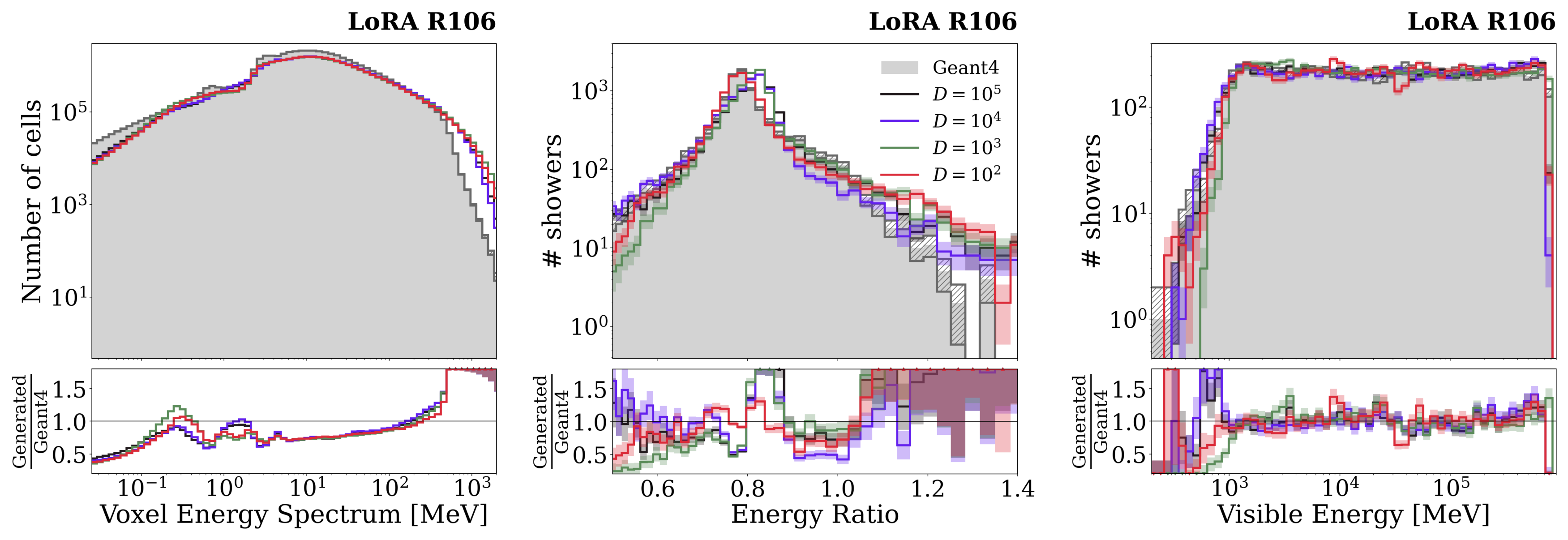}
    \end{subfigure}
    \vspace{0.2cm}
    \begin{subfigure}[b]{0.94\linewidth}
        \centering
        \includegraphics[width=\linewidth]{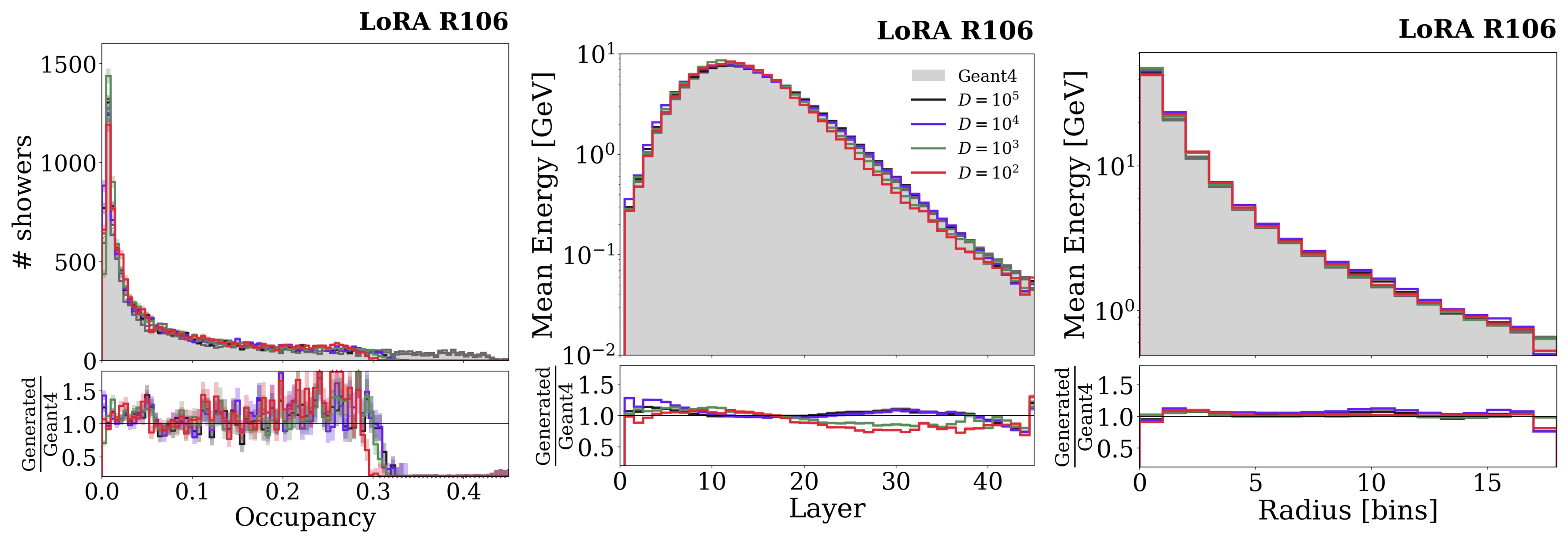}
    \end{subfigure}
    \caption{\textsc{Geant4} vs generated showers for parameter-efficient methods at training sizes $D$. 
    A comprehensive distribution analysis of generated showers for \textsc{LoRA R8} (first two rows), and \textsc{LoRA R106} (last two rows).
    All histograms from $10,000$ events with energies logarithmically distributed from 1 to 1000 GeV. 
    Bottom panels show \textsc{Geant4} ratios with statistical uncertainties.
    The error band represents the statistical uncertainty in each bin.}
    \label{fig:comparison_lora_variants}
\end{figure}

Figure \ref{fig:comparison_peft_methods} and \ref{fig:comparison_lora_variants} present detailed distribution comparisons between \textsc{Geant4} reference data and generated showers for all PEFT methods at various training dataset sizes.
These histograms reveal method-specific strengths and weaknesses: \textsc{BitFit} maintains stable energy spectrum reconstruction across all scales, \textsc{Top2} fine-tuning shows particularly good longitudinal profile modeling, while \textsc{LoRA} variants exhibit systematic biases in occupancy and radial distributions that persist even with increased training data.

\subsection{KL evaluation}
\label{app:kl_evaluation}
\begin{figure}[htbp]
    \centering
    \begin{subfigure}[b]{0.8\linewidth}
        \centering
        \includegraphics[width=\linewidth]{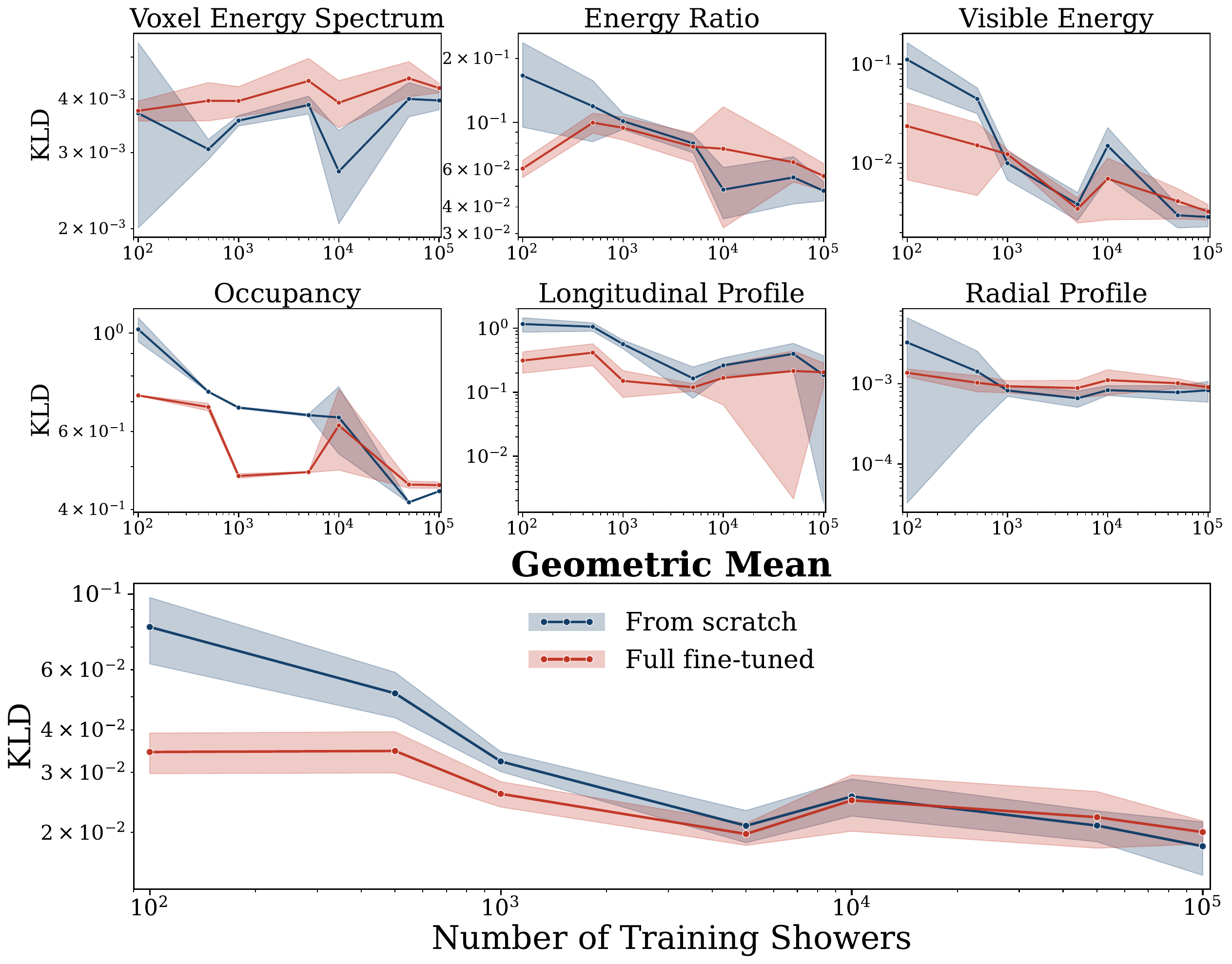}
    \end{subfigure}
    \vspace{0.25cm} 
    \begin{subfigure}[b]{0.8\linewidth}
        \centering
        \includegraphics[width=\linewidth]{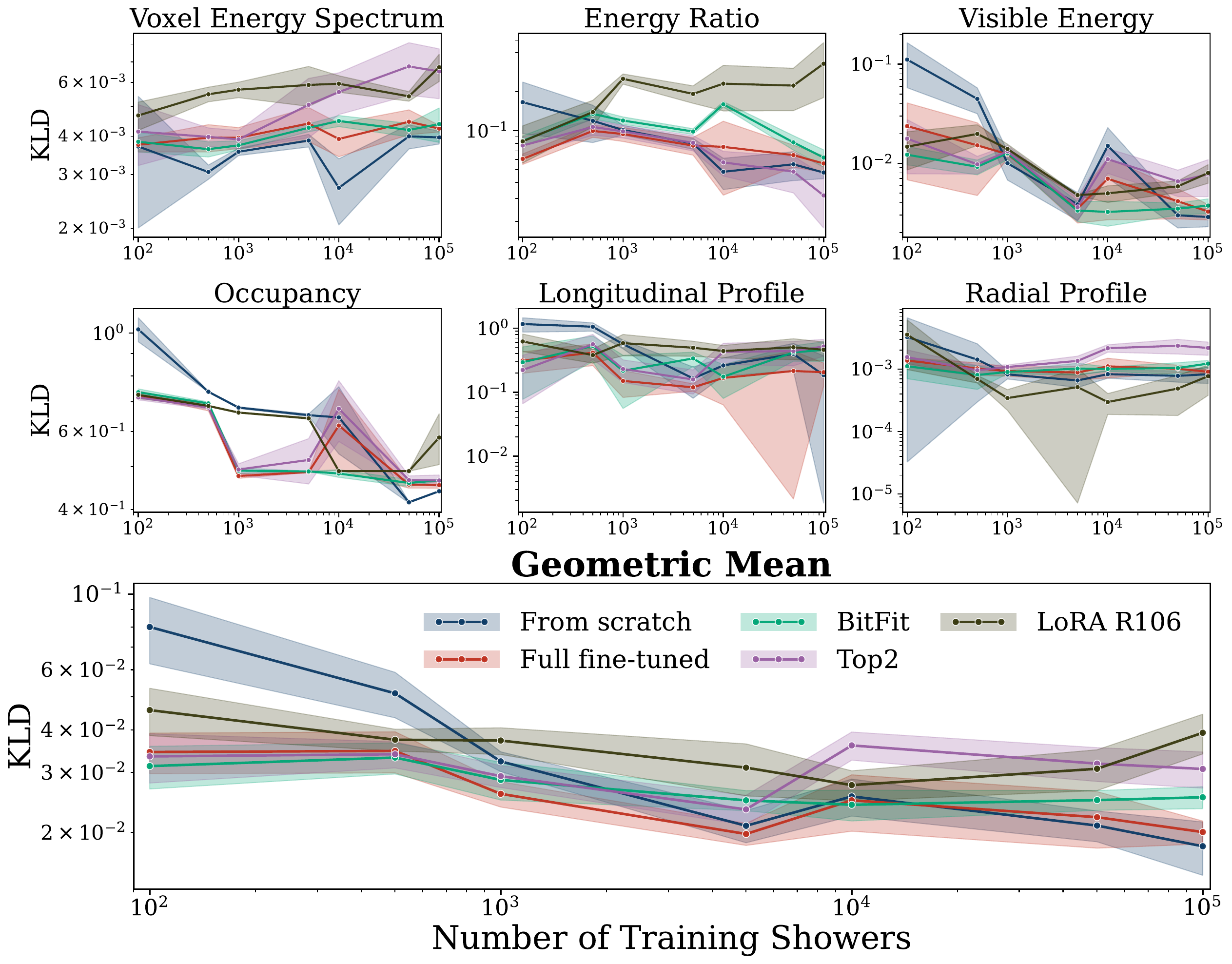}
    \end{subfigure}
    \caption{Kullback-Leibler divergence evaluation across training strategies. \textbf{Top:} \textsc{From-scratch} versus full fine-tuning comparison. \textbf{Bottom:} Complete PEFT comparison including \textsc{BitFit} (green), \textsc{Top2} (purple), and \textsc{LoRA R106} (brown). The energy ratio anomaly at $10^4$ samples is visible in both panels. Error bands represent the standard error across five random seeds.}
    \label{fig:combined_fig}
\end{figure}

To verify that our conclusions are robust to metric choice, we repeat all evaluations using the Kullback-Leibler divergence. 
Figure \ref{fig:combined_fig} presents these results across all training strategies and observables.

The KL metric confirms our main findings while revealing additional insights. Transfer learning provides consistent benefits in low-data regimes, with KL divergence reducing by 35-50\% compared to training from scratch. The energy ratio anomaly at $10^4$ samples appears in both evaluation panels, confirming this is specific to the fine-tuning pathway rather than a metric artifact. 

Among PEFT methods, \textsc{BitFit} remains most effective, closely tracking full fine-tuning performance across most observables. The exception is the voxel energy spectrum, where all PEFT methods show degradation, a pattern amplified by KL's sensitivity to distribution tails.
The logarithmic scale variations across observables reflect their different intrinsic complexities, with KL showing larger relative differences between methods than the Wasserstein distance.

\newpage
\section{Additional Experiments on Low-Rank Matrices}
\label{app:lora}

We present additional results from our investigation into the low-rank update matrices.
The results presented in this appendix show significant variability across different rank choices and dataset sizes, without clear monotonic trends.
This instability likely reflects the fundamental mismatch between \textsc{LoRA}'s low-rank assumption and the high-dimensional nature of shower physics events.
We include these results for completeness and to inform future investigations, while acknowledging that no clear optimal rank emerges from this analysis.

\subsection{Effect of the rank \texorpdfstring{$r$}{r} on the downstream task}
\label{app:rank_study}

Using the \textsc{CaloChallenge} as an example, we report the WD and KL metrics achieved by different choices of the rank $r$ after training the best number of steps. 

\begin{table*}[htbp]
\centering
\caption{Study of the $r$ parameter with WD evaluation metric.}
\label{tab:rank_wd}
\small
\begin{tabular}{@{}l|r|cccc|c@{}}
\toprule
\multirow{2}{*}{\textbf{Method}} & \multirow{2}{*}{\textbf{\makecell{\# Trainable\\Parameters}} }
& \multicolumn{4}{c|}{\textbf{Training Dataset Size}} & \multirow{2}{*}{\textbf{Mean}}  \\
& & 10$^2$ & 10$^3$ & 10$^4$ & 10$^5$ \\
\midrule
LoRA R1 & 2.67K    & 0.200 & 0.160 & 0.120 & 0.148 & 0.157 \\
LoRA R2 & 5.14K    & 0.185 & 0.173 & 0.105 & 0.153 & 0.154 \\
LoRA R4 & 10.27K   & 0.178 & 0.148 & 0.103 & 0.142 & 0.143 \\
LoRA R8 & 20.54K   & 0.132 & 0.170 & \textbf{0.097} & 0.139 & 0.135 \\
LoRA R16 & 41.10K   & 0.178 & 0.168 & 0.122 & 0.148 & 0.154 \\
LoRA R32 & 82.18K  & 0.153 & 0.158 & 0.261 & 0.148 & 0.180 \\
LoRA R48 & 123.26K  & 0.145 & 0.154 & 0.118 & 0.131 & 0.137 \\
LoRA R64 & 164.35K  & 0.149 & 0.197 & 0.134 & 0.152 & 0.158 \\
LoRA R106 & 272.21K  & \textbf{0.102} & 0.148 & 0.104 & \textbf{0.123} & \textbf{0.119} \\
LoRA R204 & 523.87K  & 0.110 & \textbf{0.128} & 0.109 & 0.146 & 0.123 \\
\bottomrule
\end{tabular}
\end{table*}

\begin{table*}[htbp]
\centering
\caption{Study of the $r$ parameter with KL evaluation metric.}
\label{tab:rank_kl}
\small
\begin{tabular}{@{}l|r|cccc|c@{}}
\toprule
\multirow{2}{*}{\textbf{Method}} & \multirow{2}{*}{\textbf{\makecell{\# Trainable\\Parameters}} }
& \multicolumn{4}{c|}{\textbf{Training Dataset Size}} & \multirow{2}{*}{\textbf{Mean}}  \\
& & 10$^2$ & 10$^3$ & 10$^4$ & 10$^5$ \\
\midrule
LoRA 1 & 2.67K & 0.216 & 0.167 & 0.226 & 0.175 & 0.196 \\
LoRA 2 & 5.14K & 0.292 & 0.277 & 0.241 & 0.229 & 0.260 \\
LoRA 4 & 10.27K & 0.359 & 0.189 & \textbf{0.114} & \textbf{0.159} & 0.205 \\
LoRA 8 & 20.54K & 0.246 & 0.215 & 0.140 & 0.193 & 0.199 \\
LoRA 16 & 41.10K & 0.275 & 0.308 & 0.158 & 0.198 & 0.235 \\
LoRA 32 & 82.18K  & 0.217 & 0.197 & 0.287 & 0.178 & 0.220\\
LoRA 48 & 123.26K  & 0.214 & \textbf{0.164} & 0.190 & 0.193 & \textbf{0.190} \\
LoRA 64 & 164.35K  & 0.315 & 0.277 & 0.220 & 0.216 & 0.257 \\
LoRA 106 & 272.21K  & 0.209 & 0.241 & 0.188 & 0.224 & 0.215 \\
LoRA 204 & 523.87K  & \textbf{0.161} & 0.223 & 0.197 & 0.198 & 0.195 \\
\bottomrule
\end{tabular}
\end{table*}

We present our results in Table \ref{tab:rank_wd} and \ref{tab:rank_kl}.
The optimal rank for \textsc{CaloClouds} is between $48$ and $106$, depending on the metric used.
Note that the relationship between model size and the optimal rank for adaptation is still an open question.

The lack of clear trends supports our main finding that \textsc{LoRA} is poorly suited for this application. 
The optimal rank appears to vary unpredictably with dataset size, suggesting that the weight updates required for shower physics adaptation do not naturally decompose into low-rank structures.

\subsection{Understanding \textsc{LoRA} Limitations through Post-Hoc Weight Analysis}
\label{app:lora_svd}

To investigate why \textsc{LoRA} underperforms in the point cloud generation task, we conduct a post-hoc analysis of weight differences from successful full fine-tuning.\footnote{This analysis examines weight differences post-hoc to understand the rank requirements of successful fine-tuning. We emphasize that this provides theoretical insight into transformation complexity but represents optimistic bounds that do not capture actual \textsc{LoRA} training dynamics, since it does not capture inter-layer dependencies or gradient dynamics during actual \textsc{LoRA} training. 
The reconstruction errors presented are thus lower bounds; actual \textsc{LoRA} training faces additional challenges from joint optimization across layers typically yields higher errors due to coupling effects~\cite{aghajanyan-etal-2021-intrinsic}.}
 This inverse \textsc{LoRA} decomposition analyses the actual weight updates from full fine-tuning to determine whether these transformations are inherently high rank, providing theoretical grounding for \textsc{LoRA}'s limited effectiveness.

Given a pre-trained model with weights $W_{\text{pre}} \in \mathbb{R}^{m \times n}$ and a fully fine-tuned model with weights $W_{\text{ft}}$, we compute the weight update as:
\begin{equation}
\Delta W = W_{\text{ft}} - W_{\text{pre}} \in \mathbb{R}^{m \times n}.
\end{equation}

The Singular Value Decomposition (SVD) factorises this matrix as
\begin{equation}
\Delta W = U \Sigma V^{\top} = \sum_{i=1}^{\rho} \sigma_i u_i v_i^{\top},
\end{equation}
where $\rho = \min(m,n)$ is the maximum possible rank of $\Delta W$. For a matrix of dimension $m \times n$, the rank cannot exceed the smaller dimension; for instance, a $512 \times 256$ matrix has at most rank 256.

According to the Eckart-Young-Mirsky theorem~\cite{eckart1936approximation}, the optimal rank $r$ approximation minimizing Frobenius norm error is:
\begin{equation}
\Delta W_r = \sum_{i=1}^{r} \sigma_i u_i v_i^{\top}.
\end{equation}

The reconstruction error $\epsilon_r$ is defined as the relative Frobenius norm:
\begin{equation}
\varepsilon_r = \frac{\|\Delta W - \Delta W_r\|_F}{\|\Delta W\|_F}.
\end{equation}

Using the orthogonality properties of SVD, this can be expressed in terms of singular values. Since $\|\Delta W\|_F^2 = \sum_{i=1}^{\rho} \sigma_i^2$ and the residual $\Delta W - \Delta W_r$ contains only the truncated singular values, we have $\|\Delta W - \Delta W_r\|_F^2 = \sum_{i=r+1}^{\rho} \sigma_i^2$; therefore
\begin{equation}
\varepsilon_r = \left(\frac{\sum_{i=r+1}^{\rho} \sigma_i^2}{\sum_{i=1}^{\rho} \sigma_i^2} \right)^{1/2}.
\end{equation}

Table~\ref{tab:layer_reconstruction} presents the layer wise theoretical minimum reconstruction errors. The results immediately reveal a striking pattern: boundary layers (0 and 5) achieve perfect reconstruction at their maximum rank of 4, while internal layers exhibit severe approximation errors even at rank 106. This heterogeneity poses a fundamental challenge for uniform rank allocation strategies.

\begin{table}[htbp]
\centering
\caption{Layer-wise theoretical minimum reconstruction errors for LoRA approximations of full fine-tuning updates. Analysis performed independently per layer without inter-layer coupling.}
\label{tab:layer_reconstruction}
\small
\begin{tabular}{@{}l|c|c|rr|rr|r@{}}
\toprule
\textbf{Layer} & \textbf{Shape} & \textbf{Max Rank} & \multicolumn{2}{c|}{\textbf{Rank 8}} & \multicolumn{2}{c|}{\textbf{Rank 106}} & \textbf{95\% Energy} \\
& $(m \times n)$ & $\rho = \min(m,n)$ & \textbf{$\epsilon_8$ (\%)} & \textbf{Quality} & \textbf{$\epsilon_{106}$ (\%)} & \textbf{Quality} & \textbf{Rank} \\
\midrule
Layer 0 & $128 \times 4$ & 4 & $<0.01^*$ & Saturated & $<0.01^*$ & Saturated & 4 \\
Layer 1 & $256 \times 128$ & 128 & 65.1 & Poor & 3.5 & Acceptable & 47 \\
Layer 2 & $512 \times 256$ & 256 & 74.8 & Poor & 19.9 & Poor & 97 \\
Layer 3 & $256 \times 512$ & 256 & 74.9 & Poor & 22.3 & Poor & 106 \\
Layer 4 & $128 \times 256$ & 128 & 69.4 & Poor & 4.7 & Acceptable & 57 \\
Layer 5 & $4 \times 128$ & 4 & $<0.01^*$ & Saturated & $<0.01^*$ & Saturated & 4 \\
\bottomrule
\end{tabular}
\vspace{1mm}
\footnotesize{$^*$Rank exceeds maximum possible rank; exact reconstruction .}
\end{table}

\begin{figure}[htbp]
\centering
\includegraphics[width=0.92\linewidth]{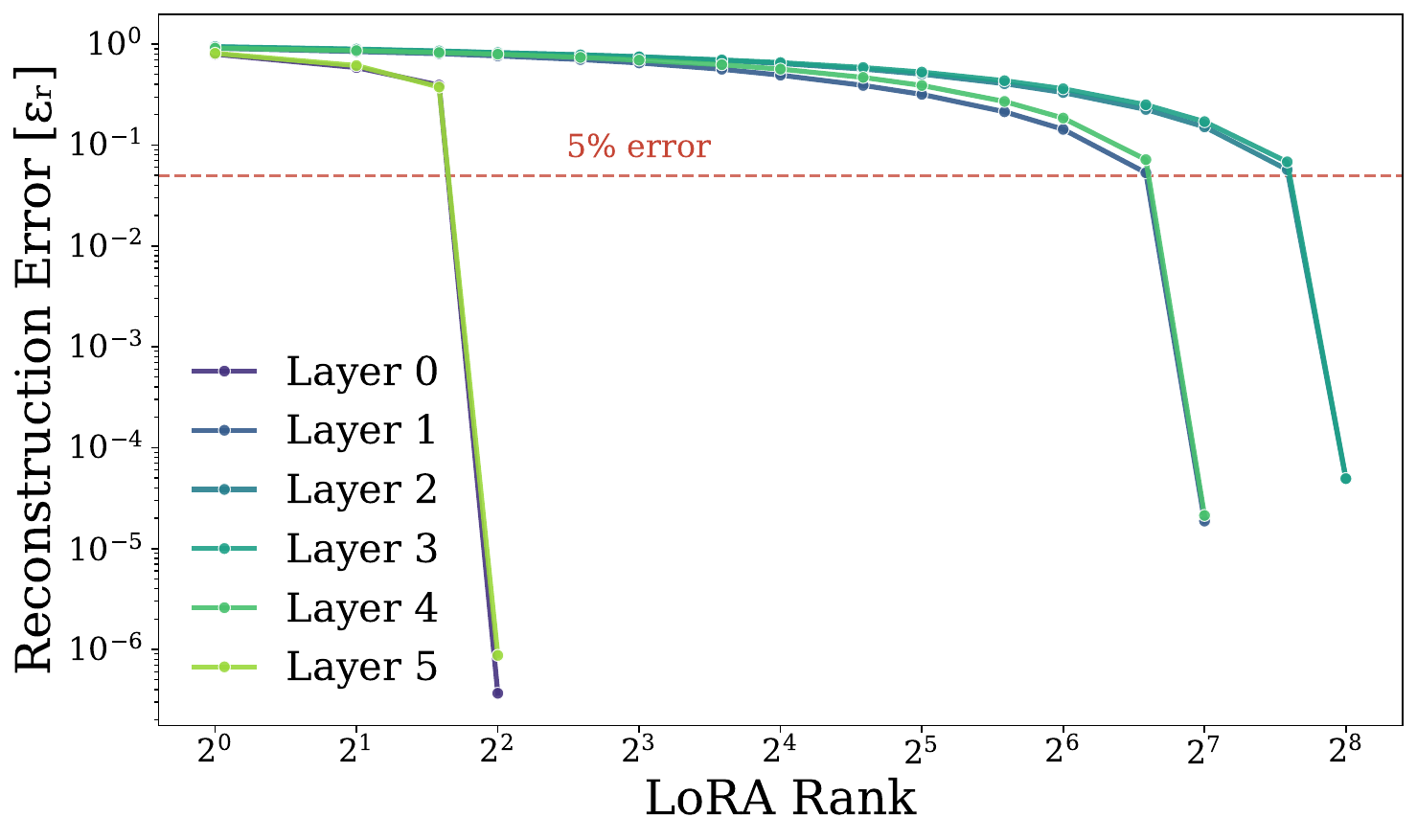}
\caption{Per-layer theoretical minimum reconstruction error $\epsilon_r$ for LoRA approximations. Analysis performed on individual layers without inter-layer coupling. Actual LoRA training would yield higher errors due to joint optimisation constraints and gradient coupling across layers.}
\label{fig:rank_decomposition}
\end{figure}

\begin{figure}[htbp]
\centering
\includegraphics[width=0.92\linewidth]{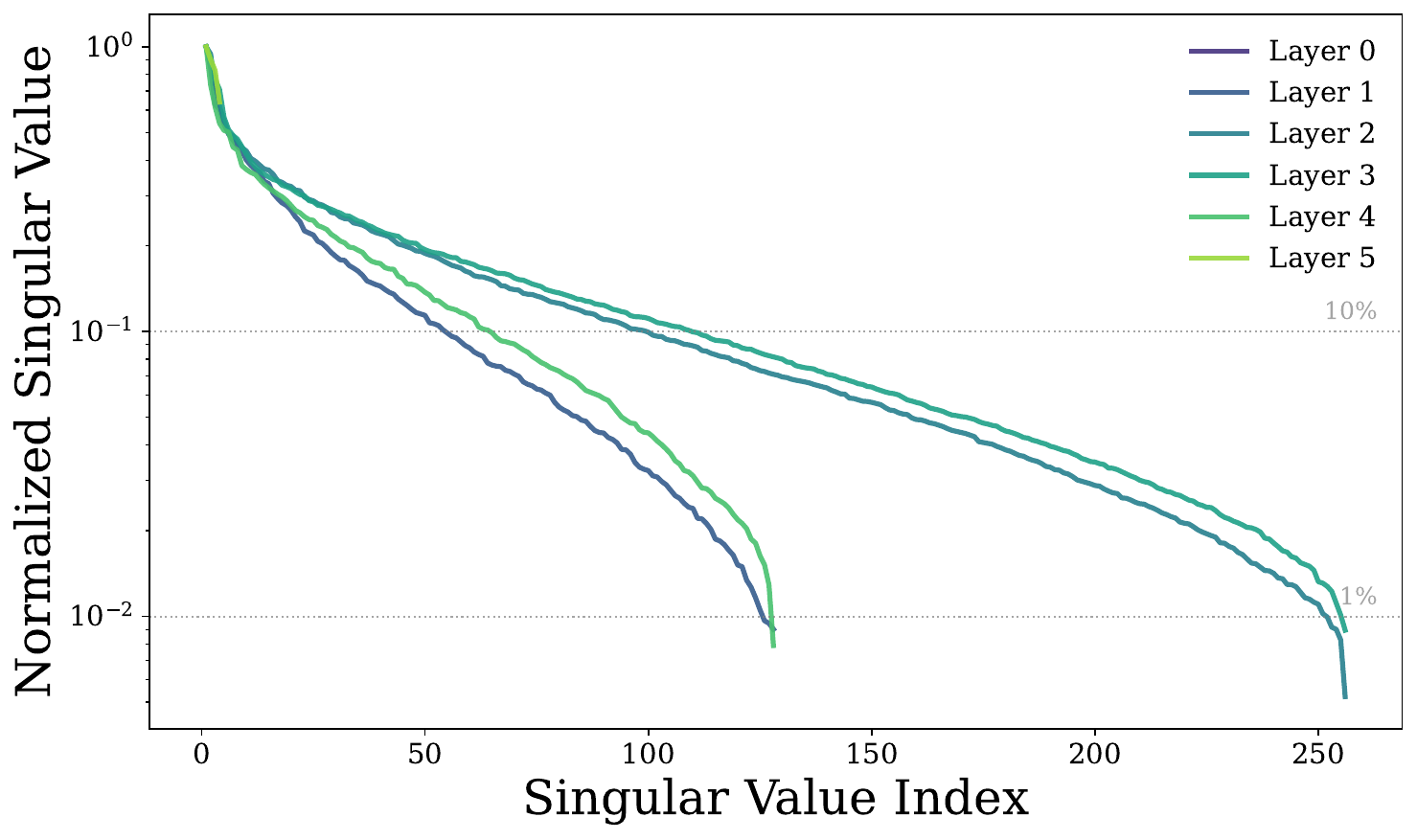}
\caption{Normalized singular value spectrum ($\tilde{\sigma}_i = \sigma_i/\sigma_1$) of weight updates from full fine-tuning. The slow decay in layers 2 and 3 indicates high intrinsic dimensionality incompatible with low rank approximation, while the sharp drops in layers 0 and 5 reflect their rank 4 constraint.}
\label{fig:svdecay}
\end{figure}

Figure~\ref{fig:rank_decomposition} illustrates how reconstruction error varies dramatically across layers as rank increases. Layers 2 and 3, which encode the most complex transformations, show particularly slow error reduction, remaining above 20\% error even at rank 106. The singular value spectrum in Figure~\ref{fig:svdecay} provides deeper insight into this phenomenon. The normalized singular values reveal that layers 2 and 3 maintain significant magnitude even at high indices, with values staying above 1\% of the maximum past index 250. This slow decay indicates these transformations span nearly the full parameter space rather than concentrating in a low dimensional subspace.

The singular value analysis reveals critical limitations for physics applications. Achieving 95\% energy capture requires ranks of 47, 97, 106, and 57 for layers 1 through 4 respectively.
The compression ratios of only 2.4 to 2.6 times for critical layers contrast sharply with the 10 to 100 times compression achieved in NLP tasks where LoRA succeeds~\cite{hu2021loralowrankadaptationlarge}.


These theoretical bounds suggest that successful adaptation requires higher ranks than commonly used in NLP applications.
While this analysis doesn't capture full training dynamics, it provides useful insight into why LoRA underperforms in our experiments and may guide future development of physics-specific PEFT methods.
These findings suggest that the low-rank assumption underlying LoRA may be less suitable for physics transformations than for language tasks. While not conclusive, this analysis provides a starting point for understanding PEFT limitations in scientific applications and motivates exploration of alternative approaches that can accommodate heterogeneous complexity across network layers.

\section{Geometric Mean and Error Propagation}
\label{app:geom_mean}

For aggregating performance metrics across different observables with disparate scales, we employ a weighted geometric mean computed in logarithmic space. Given $n$ metrics with values $\{y_i\}_{i=1}^n$, standard deviations $\{\sigma_i\}_{i=1}^n$, and weights $\{w_i\}_{i=1}^n$ (where $\sum_i w_i = 1$), the geometric mean and its uncertainty are computed as follows.

\subsection{Geometric Mean Calculation}

The weighted geometric mean is defined as:
\begin{equation}
    \bar{y}_{\text{geom}} = \prod_{i=1}^{n} y_i^{w_i} = \exp\left(\sum_{i=1}^{n} w_i \ln y_i\right).
\end{equation}

In practice, we compute this in base-10 logarithm for numerical stability:
\begin{equation}
    \bar{y}_{\text{geom}} = 10^{\bar{L}},
\end{equation}
where the mean in log-space is:
\begin{equation}
    \bar{L} = \sum_{i=1}^{n} w_i \log_{10}(y_i + \epsilon),
\end{equation}
with $\epsilon = 10^{-10}$ added to avoid numerical issues with zero values.

\subsection{Error Propagation}

The uncertainty propagation through the logarithmic transformation follows from the delta method. For a value $y_i$ with standard deviation $\sigma_i$, the uncertainty in log-space is:
\begin{equation}
    \sigma_{\log,i} = \frac{\sigma_i}{(y_i + \epsilon) \ln(10)} .
\end{equation}

The weighted variance in log-space becomes:
\begin{equation}
    \sigma^2_{\bar{L}} = \sum_{i=1}^{n} w_i^2 \sigma^2_{\log,i}.
\end{equation}

Finally, the standard deviation of the geometric mean is obtained by transforming back from log-space:
\begin{equation}
    \sigma_{\bar{y}_{\text{geom}}} = \bar{y}_{\text{geom}} \cdot \ln(10) \cdot \sigma_{\bar{L}}.
\end{equation}

This approach ensures proper handling of metrics spanning multiple orders of magnitude while maintaining mathematically consistent error propagation.

\newpage
\bibliographystyle{JHEP}  
\bibliography{biblio}     

\end{document}